\documentclass[pre, twocolumn]{revtex4}
\usepackage[dvips]{graphicx}
\usepackage{color}
\usepackage{amsmath}
\usepackage{amssymb}
\usepackage{psfrag}
\newcommand{\parfracA}[2]{\frac{\partial #1}{\partial #2}}
\newcommand{\parfracB}[3]{\frac{\partial^{#3} #1}{\partial #2^{#3}}}
\newcommand{\fH}{\bar{f}}
\newcommand{\muH}{\bar{\mu}}
\newcommand{\RinsH}{\bar{R}_{\text{ins}}}
\newcommand{\RextH}{\bar{R}_{\text{ext}}}
\newcommand{\Rins}{R_{\text{ins}}}
\newcommand{\Rext}{R_{\text{ext}}}
\newcommand{\kins}{k_{\text{ins}}}
\newcommand{\kext}{k_{\text{ext}}}
\newcommand{\tilx}{\tilde{x}}
\newcommand{\tilt}{\tilde{t}}
\newcommand{\tila}{\tilde{a}}
\newcommand{\tilmu}{\tilde{\mu}}
\newcommand{\tillambda}{\tilde{\lambda}}
\newcommand{\tilRinsH}{\tilde{R}_{\text{ins}}}
\newcommand{\tilRextH}{\tilde{R}_{\text{ext}}}
\newcommand{\tilv}{\tilde{v}}
\newcommand{\tilw}{\tilde{w}}

\begin{document}

\title{Anisotropic surface reaction limited phase
transformation~dynamics in LiFePO$_{4}$}

\author{G.~K.~Singh}
\author{M.~Z.~Bazant}
\affiliation{Department of Mathematics,\\
Massachusetts Institute of Technology,\\
Cambridge, MA 02139, USA}
\author{G.~Ceder}
\affiliation{Department of Materials Science and Engineering,\\
Massachusetts Institute of Technology,\\
Cambridge, MA 02139, USA}

\date{\today}

\begin{abstract}
A general continuum theory is developed for ion intercalation dynamics
in a single crystal of a rechargeable battery cathode.  It is based on
an existing phase-field formulation of the bulk free energy and
incorporates two crucial effects: (i) anisotropic ionic mobility in the
crystal and (ii) surface reactions governing the flux of ions across the
electrode/electrolyte interface, depending on the local free energy
difference.  Although the phase boundary can form a classical diffusive
``shrinking core'' when the dynamics is bulk-transport-limited, the
theory also predicts a new regime of surface-reaction-limited (SRL)
dynamics, where the phase boundary extends from surface to surface along
planes of fast ionic diffusion, consistent with recent experiments on
LiFePO$_4$.  In the SRL regime, the theory produces a fundamentally new
equation for phase transformation dynamics, which admits traveling-wave
solutions.  Rather than forming a shrinking core of untransformed
material, the phase boundary advances by filling (or emptying)
successive channels of fast diffusion in the crystal.  By considering
the random nucleation of SRL phase-transformation waves, the theory
predicts a very different picture of charge/discharge dynamics from the
classical diffusion-limited model, which could affect the interpretation
of experimental data for LiFePO$_{4}$.
\end{abstract}

\maketitle

\section{\label{sec:introduction}Introduction}

LiFePO$_{4}$ is widely considered to be a promising cathode material for
Li-ion rechargeable batteries.  The high practical capacity and
reasonable operating voltage of the material, along with its nontoxicity
and potential low cost, make it well-suited for large-scale battery
applications \cite{padhi97, yamada01, chung02}.  Unlike many other
cathode materials that increase their Li concentration in a continuous
solid solution, Li$_{x}$FePO$_{4}$ only exists for $x \approx 0$ and $x
\approx 1$ \cite{delacourt05} and charges or discharges Li by changing
the fraction of phase with $x \approx 0$ and $x \approx 1$.

\begin{figure}
\includegraphics[scale=1]{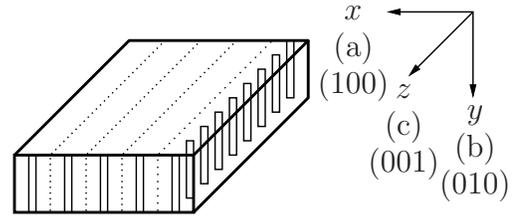}
\caption{\label{fig:crystal}Schematic of plate-like single crystals of
LiFePO$_{4}$.  Li is confined to 1D channels in the $y$ direction, and
channels are stacked in layers parallel to the $yz$ plane, indicated by
the dotted lines.  Typical dimensions of single crystals are $2 \times
0.2 \times 4$ $\mu$m in the $x$, $y$, $z$ dimensions, respectively
\cite{chen06}.  For each direction, the corresponding space group
\textit{Pnma} axis and Miller index are shown in parentheses.}
\end{figure}

The transport and phase separation properties of LiFePO$_{4}$ have been
studied extensively by atomistic simulations \cite{morgan04, ouyang04,
islam05, zhou06}.  First-principles calculations have shown that Li
diffusion in the bulk FePO$_{4}$ crystal is highly anisotropic
\cite{morgan04, ouyang04}.  Li is essentially constrained to 1D channels
in the (010) direction arranged in layers that form the crystal, as
depicted in Fig.~\ref{fig:crystal}.  The lattice mismatch at the
FePO$_{4}$/LiFePO$_{4}$ phase boundary is significant (5\% in the $x$
direction, shown in Fig.~\ref{fig:crystal}), and recent work has
investigated the differences in elastic properties between the lithiated
and unlithiated material \cite{maxisch06b}.  Atomistic simulations have
also suggested that electrons in the crystal may diffuse as small,
localized polarons confined to planes parallel to the Li channels
\cite{xu04, zhou04, maxisch06a}.

Recent experiments have confirmed the anisotropic transport and phase
separation of Li in single crystal LiFePO$_{4}$ \cite{chen06, laffont06,
amin07}.  Moreover, detailed microscopy in these studies has revealed
that the FePO$_{4}$/LiFePO$_{4}$ phase boundary is a well-defined
interface that extends through the bulk crystal to the surface.  In
experiments, the phase boundary has a characteristic width of several
nanometers on the surface \cite{laffont06}, although this width is
probably broadened by experimental resolution, and Li insertion and
extraction seem to be concentrated in this region, with negligible
transfer occurring in either the FePO$_{4}$ or LiFePO$_{4}$ phases.
Notably, the phase boundary moves orthogonally to the direction of the
surface flux, indicating that as Li insertion (extraction) proceeds,
layers of the 1D channels are progressively filled (emptied).  The
observance of surface cracks and their alignment with the phase boundary
\cite{chen06} also reinforces the view that the FePO$_{4}$/LiFePO$_{4}$
lattice mismatch plays an important role in the electrochemical function
of the material, as may the associated stress field \cite{meethong07}.

In light of the understanding gained from these atomistic and
experimental studies, the continuum theory of transport and phase
separation in LiFePO$_{4}$ merits renewed attention.  The prevailing
``shrinking core'' model can be traced to a qualitative picture
accompanying the first experimental demonstration of the material as an
intercalation electrode \cite{padhi97}.  This model assumes a growing
shell of one phase surrounding a shrinking core of the other phase, with
the shell and core phases determined by the direction of the net Li
flux: a LiFePO$_{4}$ shell surrounds an FePO$_{4}$ core during Li
insertion (battery discharging); an FePO$_{4}$ shell surrounds a
LiFePO$_{4}$ core during Li extraction (battery charging).  It is
important to note that the boundary between the shell and core phases is
entirely contained within the bulk of the material and moves parallel to
the direction of the Li flux, in contrast to the observations of the
experiments cited above.  

The current state of mathematical modeling of ion intercalation is based
on the shrinking-core concept with some further simplifying assumptions.
In earlier work, a simplified version of the model was mathematically
formulated by Srinivasan and Newman \cite{srinivasan04} and incorporated
into an existing theory for transport in composite cathodes
\cite{doyle93}.  In their model, FePO$_{4}$ is treated as a continuous,
isotropic material, and Li is inserted and extracted uniformly over the
surface of a spherical FePO$_{4}$ particle.  The phase boundary is
defined as where the compositions
Li$_{\epsilon}$FePO$_{4}$/Li$_{1-\epsilon}$FePO$_{4}$ coexist, with
$\epsilon \ll 1$ specifying the equilibrium composition between the
Li-poor and Li-rich phases, and no nucleation constraints are included.
Only Li transport in the shell is considered and modeled by an
isotropic, constant diffusivity diffusion equation, while the velocity
of the phase boundary is prescribed by a mass balance across the
boundary.  Thus, for Li insertion (extraction), diffusion in the growing
shell occurs between the surface Li concentration and $1-\epsilon$
($\epsilon$).  The value of $\epsilon$ is set as a parameter in the
numerical solution of the model.

In this paper, we propose a more general continuum theory for ionic
transport and phase separation in single-crystal rechargeable battery
materials, focusing on the special case of LiFePO$_{4}$.  Our theory
accounts for anisotropic Li diffusion in the bulk as well as the
formation and dynamics of the FePO$_{4}$/LiFePO$_{4}$ phase boundary,
driven by surface reactions at the electrolyte/electrode interface.  We
utilize an existing phase-field model for the free energy of the system
to calculate the Li chemical potential \cite{han04}; the bulk transport
equation and surface reaction rates for Li are then derived in terms of
this chemical potential.  The phase-field approach provides a sound
thermodynamic basis for studying the system and also directly connects
our theory to atomistic modeling, since first-principles simulations can
accurately compute the Li chemical potential \cite{dedompablo02,
vanderven04}.

\begin{figure*}
\includegraphics[scale=1]{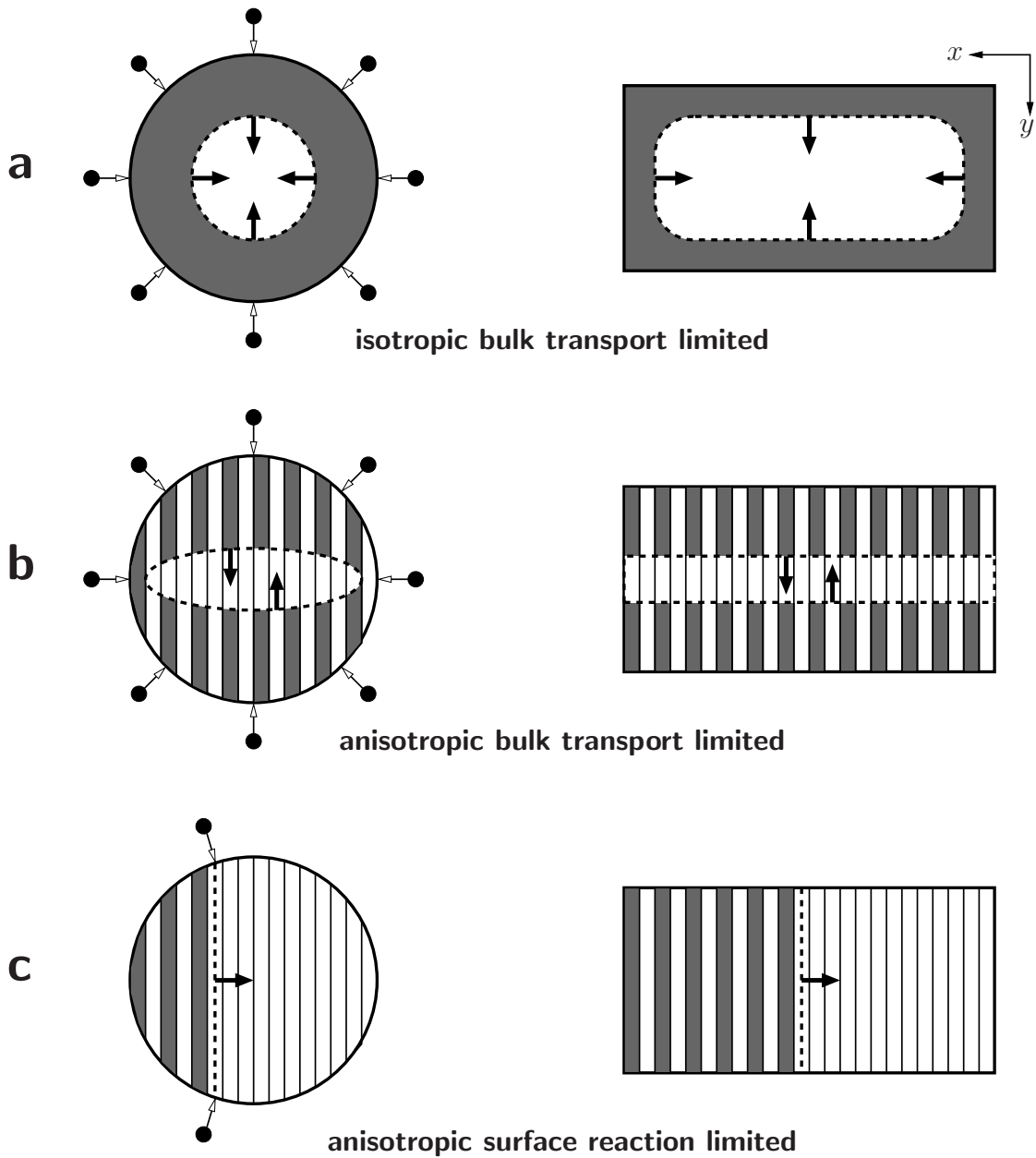}
\caption{\label{fig:models} Transport models obtained in different
limits of the characteristic timescales for bulk diffusion and surface
reactions.  Figures show $xy$ cross sections of spherical and plate-like
single crystals during Li insertion, after phase nucleation has
occurred.  Lithiated portions of the crystal are shaded, and points
outside particles represent flux of Li ions across the
electrode/electrolyte interface (shown only for spherical particles).
The FePO$_{4}$/LiFePO$_{4}$ phase boundary is denoted by the dashed
line, and arrows indicate movement of the boundary as insertion
proceeds.  (a) Isotropic bulk transport limited.  (b) Anisotropic bulk
transport limited.  (c) Anisotropic surface reaction limited.}
\end{figure*}

We first develop a general model that encapsulates various transport
regimes in different limits of the characteristic timescales for bulk
diffusion and surface reactions, as presented in Fig.~\ref{fig:models}.
The characteristic timescales for bulk diffusion $t_{D}$ and surface
reactions $t_{R}$ are
\begin{equation}
t_{D} = \frac{L^{2}}{D}, \quad t_{R} = \frac{1}{k},
\end{equation}
where $L$ is the lengthscale over which diffusion occurs, $D$ is the
diffusivity, and $k$ is the surface reaction rate.  The dimensionless
ratio of these timescales is the Damkohler number
\begin{equation}
Da = \frac{t_{D}}{t_{R}} = \frac{k L^{2}}{D}.
\end{equation}
Therefore, an isotropic bulk transport limited (BTL) process, where bulk
diffusion in all directions is much slower than surface reactions, is
characterized by
\begin{equation}
Da \gg 1.
\end{equation}
In this regime, the phase boundary is entirely contained within the
material and moves along the direction of the Li flux, as shown in
Fig.~\ref{fig:models}a.  Anisotropic BTL phase transformation in
LiFePO$_{4}$ is depicted in Fig.~\ref{fig:models}b.  Bulk diffusion in
the $x$ and $z$ directions is negligible, and bulk diffusion in the $y$
direction is much slower than surface reactions.  Consequently,
\begin{equation}
Da_{x}, Da_{z} \gg Da_{y} \gg 1,
\end{equation}
where $Da_{x} = k L^{2} / D_{x}$ with $D_{x}$ the diffusivity in the $x$
direction, and similarly for $Da_{y}$ and $Da_{z}$.  In this regime, the
phase boundary is still contained within the bulk, but Li is confined to
1D channels in the $y$ direction.  The anisotropic BTL model is a
generalization of the shrinking core model developed by Srinivasan and
Newman \cite{srinivasan04}.  However, motivated by the experiments and
simulations described above, we focus on a different transport regime
where surface reactions are much slower than diffusion in the $y$
direction but much faster than diffusion in the $x$ and $z$ directions.
Thus, we study the limit
\begin{equation}
Da_{x}, Da_{z} \gg 1 \gg Da_{y},
\end{equation}
illustrated in Fig.~\ref{fig:models}c.  The phase boundary in this
regime extends through the bulk of the material to the $xz$ surfaces.
Moreover, the surface reaction rates of Li transfer are concentrated at
the phase boundary such that each channel is almost completely lithiated
(delithiated) before insertion (extraction) progresses to an adjacent
channel.  In such an anisotropic surface reaction limited (SRL) process,
our general model reduces to a fundamentally new equation governing the
transport and phase separation dynamics.  Analysis and numerical
solutions of this equation show that it qualitatively reproduces the
features of the LiFePO$_{4}$ system observed in experiments.

In this initial effort, we neglect the possibility of charge separation
and assume that electrons are freely available in the material to
compensate the charge of Li$^{+}$, although the presence of multiple
diffusing and migrating species (interacting through an electrostatic
potential) can be modeled as an extension of the general framework we
present.  We also avoid an explicit treatment of stress around the
FePO$_{4}$/LiFePO$_{4}$ interface; instead, we note that a term in the
phase field formulation may serve as an approximation of the energy due
to the lattice mismatch.

\section{\label{sec:gmodel}General Model}

\subsection{Phase field formulation}

We follow the conventional Cahn-Hilliard formulation \cite{cahn58} that
has been previously applied as a phase field model for bulk transport in
LiFePO$_{4}$ \cite{han04}.  The total free energy of the system is
expressed as a functional of the local Li concentration
\begin{equation}\label{eqn:F} 
F = \int \left[ \fH(c) + (K/2) |\nabla c|^{2} \right] d\mathbf{r}, 
\end{equation} 
where $c$ is the dimensionless, normalized Li concentration ($0 < c <
1$), $\fH(c)$ is the homogeneous free energy density, and $K$ is the
gradient energy coefficient that represents the energy penalty for
maintaining concentration gradients in the system.  The lattice mismatch
at the FePO$_{4}$/LiFePO$_{4}$ phase boundary coincides with the
concentration gradient, so the gradient penalty in \eqref{eqn:F} can be
regarded as approximating related contributions the free energy.
Phase-field models have also been developed, which include the
long-range elastic contributions to the free energy
\cite{khachaturyan83, larche85}, but here we restrict ourselves to the
simpler formulation above, to allow us to focus on other effects, namely
anisotropic transport and surface reactions. In this spirit, we also
ignore any surface contributions to the free energy, such as the tension
of the electrode/electrolyte interface.

The homogeneous, bulk free energy density takes the form of a regular
solution model
\begin{equation}\label{eqn:fH}
\fH(c) = a c (1 - c) + \rho k T \left[ c \ln c + (1 - c) \ln (1 - c)
\right],
\end{equation}
where $a$ is the average energy density (in a mean field sense) of the
interaction between Li ions, $\rho$ is the number of intercalation
sites per unit volume, $k$ is Boltzmann's constant, and $T$ is the
temperature.  The first term in \eqref{eqn:fH}, the enthalpic
contribution to the free energy, promotes separation of the system to
$c = 0$ or $c = 1$, while the second term, the entropic contribution,
favors mixing of the system.  Therefore, the strength of the phase
separation is characterized by the dimensionless ratio $\rho k T / a$.
The chemical potential of Li in the FePO$_{4}$ host is calculated as
the variational derivative
\begin{align}
\mu &= \frac{\delta F}{\delta c},\\
&= \muH - K \nabla^{2} c,\label{eqn:mu}
\end{align}
where we define $\muH$ as the homogeneous chemical potential
\begin{align}
\muH &= \parfracA{\fH}{c},\\
&= a (1 - 2 c) + \rho k T \ln \left( \frac{c}{1 - c}
\right).\label{eqn:muH}
\end{align}

While in general, the two phase compositions in equilibrium across the
miscibility gap are determined by the common tangent construction, in
the symmetric free energy $\fH$ of \eqref{eqn:fH} these compositions
correspond to $\muH = 0$.  These roots cannot be found analytically from
\eqref{eqn:muH}, but asymptotic approximations in the small parameter
$\rho k T / a$ can be obtained; two term expansions are
\begin{align}
c_{-} &\sim e^{-\frac{a}{\rho k T}} \left( 1 + \frac{a}{\rho k T}
e^{-\frac{a}{\rho k T}} \right),\\
c_{+} &\sim 1 - c_{-},
\end{align}
where $c_{-}$ is the root near $c = 0$ and $c_{+}$ is the root near $c =
1$.  As may be expected, $c_{\pm}$ approach the concentration extremes
exponentially in $a/(\rho k T)$.  While nucleation may be required to
form a second phase for compositions in the miscibility gap, spontaneous
phase separation occurs when the composition is within the spinodals.
The spinodals correspond to the zeros of the curvature of the free
energy and can be determined from \eqref{eqn:fH} as
\begin{equation}
c_{\text{sp}} = \frac{1 \pm \sqrt{1 - 2 \rho k T / a}}{2}.
\end{equation}
We observe that $a > 2 \rho k T$ is required for distinct, physically
meaningful spinodal compositions.

\subsection{Anisotropic bulk transport}

As described above, Li migration in the bulk crystal is confined to 1D
channels in the (010) direction, labeled as $y$ in
Fig.~\ref{fig:models}.  Some diffusion in other directions may occur due
to defects in the crystal lattice or cracks caused by the
FePO$_{4}$/LiFePO$_{4}$ lattice mismatch, as have been observed
experimentally \cite{chen06, laffont06}.  Experiments \cite{chen06,
laffont06} have found that layers of stacked 1D channels in the $z$
direction are progressively filled or emptied as the phase boundary
moves across the layers in the $x$ direction, indicating that transport
in the $z$ direction is faster than transport in the $x$ direction.  The
anisotropic Li flux can be written as
\begin{equation}
\mathbf{j} = -c \mathbf{B} \nabla \mu,
\end{equation}
where
\begin{equation}
\mathbf{B} = \begin{pmatrix}
b_{11} & 0 & 0\\
0 & b_{22} & 0\\
0 & 0 & b_{33}
\end{pmatrix}
\end{equation}
is the mobility tensor for an orthorhombic crystal, and the indices 1,
2, 3 correspond to the $x$, $y$, $z$ directions, respectively.  The
diffusivity tensor is determined from the mobility by the Einstein
relation $\mathbf{D} = k T \mathbf{B}$. Based on the discussion above,
we expect that $b_{11} \ll b_{33} \ll b_{22}$.  We also note that
$\mathbf{D}$ can, in general, be position or concentration dependent.
In fact, for some other intercalation materials that do not phase
separate, first principles calculations have shown that the diffusivity
is a highly nonlinear function of the Li concentration, varying by
several orders of magnitude over the composition range
\cite{vanderven00}.

With the ionic fluxes thus defined, the dynamics of the concentration
profile is governed by the conservation law
\begin{equation}\label{eqn:Li}
\rho \parfracA{c}{t} + \nabla \cdot \mathbf{j} = 0,
\end{equation}
where the factor of $\rho$ is needed for dimensional consistency.

\subsection{Surface reactions}

We assume that Arrhenius kinetics govern the insertion and extraction
rates of Li across the electrode/electrolyte interface.  The activation
energies for these interfacial reactions are defined as the change in
the chemical potential of Li across the interface.  We further assume
that the chemical potential of Li in the FePO$_{4}$ host, $\mu$ given by
\eqref{eqn:mu}, is valid everywhere on the electrode surface where Li
transfer occurs, with the appropriate modification for the Laplacian
term. By also using the bulk chemical potential for ions at the crystal
surface, we are neglecting the possibility of any variation in the
chemical potential at the electrode/electrolyte interface, e.g. due to
surface orientation or surface curvature, as noted above \cite{wang07}.

With these assumptions, the local Li insertion rate is 
\begin{equation}
\Rins = \kins c_{e} e^{(\mu_{e} - \mu) / (\rho k T)},
\end{equation}
where $\kins$ is the insertion rate coefficient, and $c_{e}$ and
$\mu_{e}$ are the concentration and chemical potential of Li in the
electrolyte, respectively.  Note that since $c_{e}$ is expressed as a
dimensionless filling fraction, $\Rins$ and $\kins$ have dimensions of
inverse time.  

In a more complete battery model, $c_{e}$ and $\mu_{e}$ would be
determined by solving the appropriate transport equations for Li in the
electrolyte.  However, as we focus here on Li transport in the
electrode, we ignore variations in the electrolyte and take $c_{e}$ and
$\mu_{e}$ as constants.  Our formulation therefore describes
potentiostatic, or constant chemical equilibrium, conditions in the
electrolyte, with the interfacial transfer of Li as the rate limiting
process.  Absorbing $c_{e}$ into $\kins$ gives
\begin{align}
\Rins &= \kins e^{\alpha (\mu_{e} - \mu)},\\
&= \RinsH e^{\alpha K \nabla^{2} c},
\end{align}
where $\alpha = 1 / (\rho k T)$, and $\RinsH$ is the homogeneous
insertion rate
\begin{equation}
\RinsH = \kins \left( \frac{1 - c}{c} \right)
  e^{\alpha [\mu_{e} - a (1 - 2 c)]}.
\end{equation}
Similarly, the extraction rate is
\begin{align}
\Rext &= \kext c e^{\alpha (\mu - \mu_{e})},\\
&= \RextH e^{-\alpha K \nabla^{2} c},
\end{align}
where $\RextH$ is the homogeneous extraction rate
\begin{equation}
\RextH = \kext \left( \frac{c^{2}}{1 - c} \right)
  e^{\alpha [a (1 - 2 c) - \mu_{e}]}.
\end{equation}
In contrast to $\Rins$, the proportionality of $\Rext$ on $c$ cannot be
neglected and breaks the symmetry about $c = 1/2$.  The net rate of Li
insertion is therefore
\begin{align}
R &= \Rins - \Rext,\\
&= \RinsH e^{\alpha K \nabla^{2} c} - \RextH e^{-\alpha K \nabla^{2}
c},\\
\begin{split}
&= \kins \left( \frac{1 - c}{c} \right)
  e^{\alpha [\mu_{e} - a (1 - 2 c) + K \nabla^{2} c]}\\
&\quad - \kext \left( \frac{c^{2}}{1 - c} \right)
  e^{\alpha [a (1 - 2 c) - \mu_{e} - K \nabla^{2} c]}.
\end{split}
\end{align}

The boundary conditions for \eqref{eqn:Li} on the crystal surface
express mass conservation at the electrode/electrolyte interface,
\begin{equation}\label{eqn:surfbc}
\mathbf{n} \cdot \mathbf{j} = -\rho_{s} R,
\end{equation}
where $\mathbf{n}$ is the unit normal vector directed out of the
crystal, and $\rho_{s}$ is the number of intercalation sites per unit
area, which depends on the orientation of the surface.  Consistent
with our neglect of surface excess chemical potential, we neglect the
possibility of a surface flux density $\mathbf{j}_{s}$, whose surface
divergence $\nabla_{s} \cdot \mathbf{j}_{s}$ would appear as an extra
term on the right hand side of \eqref{eqn:surfbc}.

\section{\label{sec:smodel}Surface-reaction-limited phase
transformation}

\subsection{Depth-integrated dynamics}

We now develop a special limit of the general model that describes SRL
phase-transformation in LiFePO$_{4}$.  We assume that fast diffusion in
the $y$ oriented 1D channels rapidly equilibrates the bulk Li
concentration to the surface concentration.  A depth-averaged
concentration $\bar{c}$ on the $xz$ surface can therefore be defined as
\begin{equation}
\bar{c} = \frac{1}{L_{y}(x,z)} \int c(x, y, z, t) \,dy,
\end{equation}
where $L_{y}(x,z)$ is the depth of the crystal in the $y$ direction,
from surface to surface.  

By depth-averaging the bulk transport equation \eqref{eqn:Li}, using the
boundary condition \eqref{eqn:surfbc}, we find that the dynamics of
$\bar{c}$ are governed solely by the surface reaction rate $R$, acting
as a source term on the $xz$ surface,
\begin{equation}\label{eqn:barc}
\left( \frac{\rho L_{y}(x, z)}{2 \rho_{s}(x, z)} \right)
\parfracA{\bar{c}}{t} =  R(\bar{c}, \nabla^{2} \bar{c}),
\end{equation}
where $\rho_{s}(x, z)$ is the number of atoms per unit area, dependent
on the local orientation of the crystal surface.  As noted above, a
surface diffusion term could also be explicitly added to
\eqref{eqn:barc}, but we shall see that the reaction rate $R$ already
produces weak $xz$ diffusion due the gradient penalty $K$, which
suffices to propagate the phase boundary along the surface.

Equation \eqref{eqn:barc} describes a fundamentally different type of
phase transformation dynamics. From a mathematical point of view, the
striking feature is that the Laplacian term $\nabla^{2} \bar{c}$ appears
in a nonlinear source term, as opposed to the additive quasilinear term
in classical reaction-diffusion equations. We are not aware of any prior
study of this type of equation, so it begs mathematical analysis to
characterize its solutions.

Here, we begin this task by making some simplifying assumptions, which
allow us to highlight new nonlinear wave phenomena described by
\eqref{eqn:barc}.  As noted above, experiments indicate that layers of
1D channels along the $yz$ plane are progressively filled or emptied as
Li transfer proceeds, so it is natural to neglect concentration
variations in the $z$ direction as a first approximation, for a planar
phase boundary spanning the crystal. We also neglect depth variations,
$L_{y} =$ constant, and assume constant surface orientation, $\rho_{s}
=$ constant, which corresponds to the common case of a plate-like
crystal.

\subsection{Dimensionless formulation}

A dimensionless form of the model, suitable for mathematical analysis,
is found by scaling each variable to its natural units.  The Li
concentration is already expressed as a dimensionless filling fraction
per channel $\bar{c}$, which will depend on the dimensionless position
$\tilx = x/L$ and time $\tilt = t/\tau$.  Position is scaled to a length
$L$, which characterizes the size of the crystal surface along which the
SRL phase transformation propagates.  The natural time scale from
\eqref{eqn:barc} is
\begin{equation}
\tau = \frac{\rho L_{y}}{ 2 \rho_{s} \kins},
\end{equation}
which is the time required for the insertion reaction to fill a single
fast-diffusion channel in the crystal (from both sides).  Note that this
time scale is proportional to the depth of the crystal, $L_{y}$.

There are four dimensionless groups which govern the solution. The first
is the ratio of reaction-rate constants
\begin{equation}
\kappa = \frac{\kext}{\kins},
\end{equation}
which measures asymmetry in the extraction and insertion reaction
kinetics.  By scaling energy density to the thermal energy density $\rho
k T$, we arrive at three more dimensionless parameters:
\begin{equation}
\tila = \frac{a}{\rho k T}, \quad \tilmu_{e} = \frac{\mu_{e}}{\rho k T},
\end{equation}
and
\begin{equation}
\tillambda = \sqrt{\frac{K}{\rho k T L^{2}}} = \frac{\lambda}{L}.
\end{equation}
The latter formula makes it clear that the natural length scale for the
phase boundary thickness, set by the gradient penalty in the free
energy, is $\lambda = \sqrt{K/(\rho k T)}$.  Since $\lambda$ is an
atomic length scale (1 \AA -- 10 nm) much smaller than the crystal size
(10 nm -- 10 $\mu$m), the parameter $\tillambda$ is typically small and
lies in the range $10^{-5} < \tillambda < 1$.

\begin{figure}
\includegraphics[scale=1]{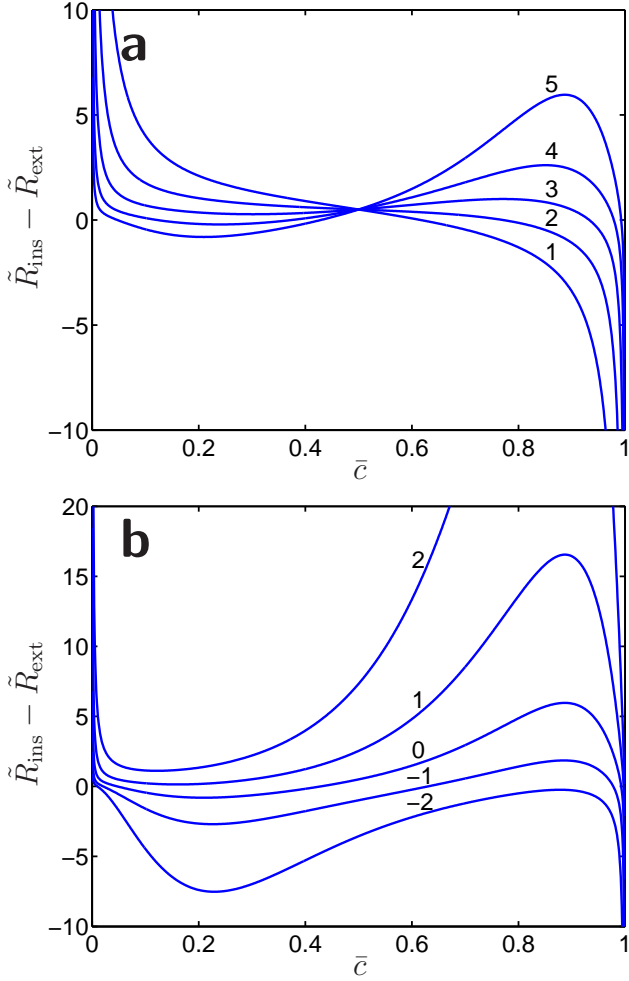}
\caption{\label{fig:RnetH} Dimensionless homogeneous net insertion rate.
(a) $\tilmu_{e} = 0$, $\kappa = 1$, $\tila = 1, \ldots, 5$ as labeled in
figure.  (b) $\tila = 5$, $\kappa = 1$, $\tilmu_{e} = -2, \ldots, 2$ as
labeled in figure.}
\end{figure}

With these scalings, the SRL phase-transformation equation
\eqref{eqn:barc} takes the dimensionless form
\begin{align}\label{eqn:nondimbarc}
\begin{split}
\parfracA{\bar{c}}{\tilt} &=
  \left( \frac{1 - \bar{c}}{\bar{c}} \right)
    e^{\tilmu_{e} - \tila (1 - 2 \bar{c}) + \tillambda^{2}
    \parfracB{\bar{c}}{\tilx}{2}}\\
&\quad - \kappa \left( \frac{\bar{c}^{2}}{1 - \bar{c}} \right)
    e^{\tila (1 - 2 \bar{c}) - \tilmu_{e} - \tillambda^{2}
    \parfracB{\bar{c}}{\tilx}{2}},
\end{split}
\end{align}
We will study solutions to this new nonlinear partial differential
equation in the following sections, but we already can gain some insight
by considering the limit of a sharp phase boundary, $\tillambda \ll 1$,
as discussed above.  Expanding \eqref{eqn:nondimbarc} for small
$\tillambda$, we obtain a reaction-diffusion equation at leading order,
\begin{equation}\label{eqn:rd}
\parfracA{\bar{c}}{\tilt} = \tillambda^{2} (\tilRinsH + \tilRextH)
\parfracB{\bar{c}}{\tilx}{2} + (\tilRinsH - \tilRextH),
\end{equation}
where $\tilRinsH$ and $\tilRextH$ are the dimensionless homogeneous
reaction rates
\begin{align}
\tilRinsH &= \left( \frac{1 - \bar{c}}{\bar{c}} \right)
  e^{\tilmu_{e} - \tila (1 - 2 \bar{c})}\\
\tilRextH &= \kappa \left( \frac{\bar{c}^{2}}{1 - \bar{c}} \right)
  e^{\tila (1 - 2 \bar{c}) - \tilmu_{e}}.
\end{align}
Thus, we see that \eqref{eqn:nondimbarc} has a direct analog to a
reaction-diffusion equation with a weak, concentration dependent
diffusivity and nonlinear source term, in the appropriate physical limit
of an atomically sharp phase boundary.  The detailed structure and
dynamics of the phase boundary, however, must be obtained by solving the
full equation \eqref{eqn:nondimbarc}.  Representative plots of the
homogeneous net insertion rate $\tilRinsH - \tilRextH$ are shown in
Fig.~\ref{fig:RnetH}.

\subsection{Traveling waves}

We seek traveling wave solutions of \eqref{eqn:nondimbarc}, which
physically correspond to waves of phase transformation, propagating
through the FePO$_{4}$ crystal with steadily translating concentration
profile.  Since $\bar{c}$ is a depth averaged concentration, these
composition waves extend through the bulk material to the surface, and
they move parallel to the surface.  Substituting the traveling-wave
ansatz 
\begin{equation}
\bar{c}(\tilx, \tilt) = g(\tilx - \tilv \tilt),
\end{equation}
into \eqref{eqn:nondimbarc}, where $\tilv$ is the unknown constant
velocity, and rewriting the resulting ordinary differential equation as
a first-order system gives
\begin{align}
g' &= h,\label{eqn:g}\\
h' &= \frac{1}{\tillambda^{2}} \ln \left[ \frac{-\tilv h + \sqrt{(\tilv
h)^{2} + 4 \kappa g}}{2 \tilRinsH(g)} \right],\label{eqn:h}
\end{align}
where $g$ and $h$ are functions of $\zeta = \tilx - \tilv \tilt$.  The
stationary points of the system are given by
\begin{align}
h &= 0,\\
\tila (1 - 2 g) + \ln \left( \frac{g^{3/2}}{1 - g} \right) + \ln \Gamma
&= 0,\label{eqn:gstnrypt}
\end{align}
where we define another dimensionless constant $\Gamma = e^{-\tilmu_{e}}
\sqrt{\kappa}$.

The solutions of \eqref{eqn:gstnrypt} correspond to the roots of the
spatially homogeneous source $\tilRinsH - \tilRextH$ and therefore
determine the equilibrium phase compositions of the system.  Explicit
expressions for the solutions are not possible, though at least one
solution must exist, since the left hand side changes sign over the
interval $0 < g < 1$.  However, phase separation into Li-poor and
Li-rich phases requires three solutions $g_{1} < g_{2} < g_{3}$, where
$g_{1}$ and $g_{3}$ are the equilibrium compositions bounding the moving
wavefront, and $g_{2}$ is an unstable intermediate composition.  We
observe that the phase compositions are independent of the dimensionless
gradient penalty $\tillambda^{2}$.  At the threshold where
\eqref{eqn:gstnrypt} transitions from one solution to three, the
solutions and extrema of the equation coincide.  The extrema occur at
the compositions
\begin{equation}\label{eqn:gextrema}
g_{\pm} = \frac{(1 + 4 \tila) \pm \sqrt{16 \tila^{2} - 40 \tila + 1}}{8
\tila},
\end{equation}
and as two distinct extrema are needed for three solutions of
\eqref{eqn:gstnrypt}, phase separation requires $\tila > 5/4 +
\sqrt{3/2} \approx 2.47$.  If the minimum $\tila$ is exceeded,
\eqref{eqn:gstnrypt} can be used to compute the rate coefficients and
electrolyte chemical potential, expressed through the combination
$\Gamma$, that will make either $g_{\pm}$ the critical composition at
threshold.  For strongly phase separating systems, such as LiFePO$_{4}$,
we find asymptotic approximations of the solutions of
\eqref{eqn:gstnrypt} in the small parameter $1/\tila$.  Two term
expansions for each root are
\begin{align}
g_{1} &\sim e^{-\tfrac{2 \tila}{3}} \left( \frac{1}{\Gamma^{2/3}} +
\frac{4 \tila e^{-\tfrac{2 \tila}{3}}}{3 \Gamma^{4/3}} \right),\\
g_{2} &\sim \frac{1}{2} - \frac{1}{2} \ln \left( \frac{\sqrt{2}}{\Gamma}
\right) \left[ \frac{1}{\tila} - \frac{5}{2 \tila^{2}} \right],\\
g_{3} &\sim 1 - e^{-\tila} \left( \Gamma + 2 \tila \Gamma^{2} e^{-\tila}
\right).
\end{align}

The existence of traveling waves and the selection of the wave velocity
in a phase separated system can be understood by a linear stability
analysis of \eqref{eqn:g}--\eqref{eqn:h} about the three stationary
points $(g_{i}, 0)$ for $i = 1, 2, 3$.  We find that $(g_{1}, 0)$ and
$(g_{3}, 0)$ are saddle points for all velocities, and $(g_{2}, 0)$ is
either a stable node or stable spiral, depending on the velocity.
Monotonic wavefronts between the equilibrium Li-poor and Li-rich phases
correspond to trajectories in the $(g, h)$ phase space that connect
$(g_{1}, 0)$ and $(g_{3}, 0)$, bypassing $(g_{2}, 0)$.  Following the
continuity arguments presented in \cite{murray02}, there is a unique
velocity $v$ such that the orientation of the eigenvectors at $(g_{1},
0)$ and $(g_{3}, 0)$ allow a single trajectory joining these points.
Thus, for a given set of parameters, all fully developed waves in a
system propagate at the same velocity.

A rigorous mathematical analysis of the traveling wave solutions of
\eqref{eqn:barc}, including their formal existence, stability, and
velocity is beyond the scope of this work.  Although analytical methods
for studying traveling waves in parabolic systems are available
\cite{volpert94}, they are usually developed for systems where there is
a diffusion term plus a source independent of derivatives of the
solution, as in \eqref{eqn:rd}.  To the best of our knowledge,
\eqref{eqn:barc} represents a different type of equation admitting
traveling wave solutions, where the curvature dependence of the source
precludes the need for an explicit diffusion term.

The nanoscale dimensions of the physical domain also complicate the
analysis of \eqref{eqn:barc}.  In other reaction diffusion equations,
such as the Fisher equation, the wave velocity is determined by assuming
an exponential decay of the leading edge of the wavefront as $\tilx
\rightarrow \infty$ \cite{murray02}.  A finite cutoff in the leading
edge is known to significantly alter the velocity \cite{brunet97}.  Such
cutoffs are present in nanoparticles of LiFePO$_{4}$, as the $xz$
surface is bounded on the scale of the wave width.  Moreover, the
nanometer wave width describes the Li concentration across only a few
atomic layers of the crystal, with each 1D channel in the layer holding
a single file of Li atoms.  Therefore, $\bar{c}$ may be discontinuous
for small particles.

\subsection{Wave propagation and nucleation}

\begin{figure}
\includegraphics[scale=1]{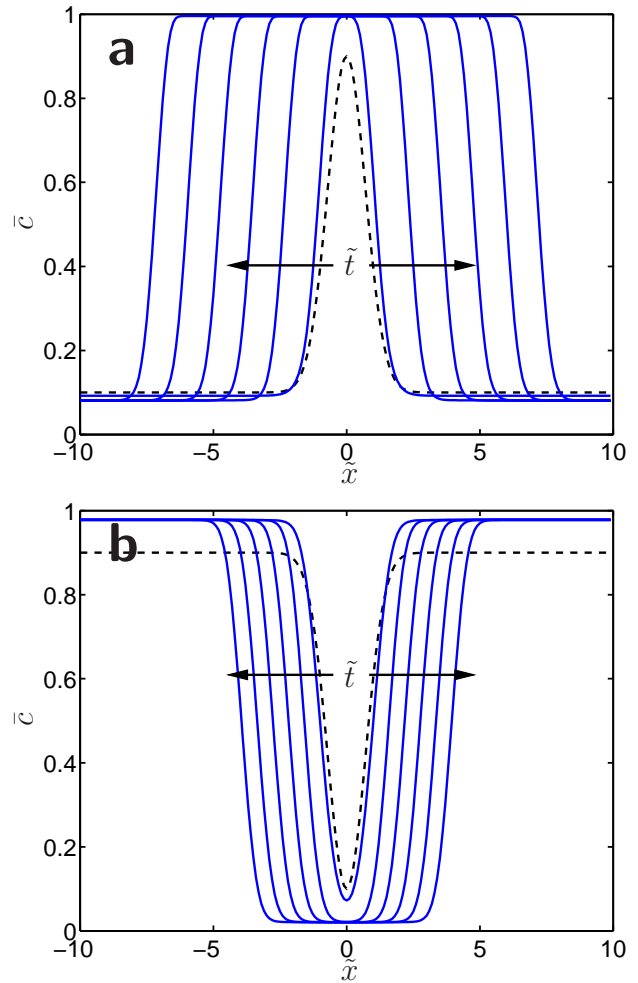}
\caption{\label{fig:twave} Numerically computed phase transformation
waves during insertion and extraction.  Dashed lines show initial
conditions; solid lines show concentration profiles at uniformly spaced
times, with arrows indicating direction of wave propagation.  For both
insertion and extraction: $\tila = 5$, $\kappa = 1$, $\tillambda = 1$.
(a) Insertion wave with $\tilmu_{e} = 0.5$, $\bar{c}(\tilx, 0) = 0.1 +
0.8 \exp(-\tilx^{2})$.  (b) Extraction wave with $\tilmu_{e} = -1$,
$\bar{c}(\tilx, 0) = 0.9 - 0.8 \exp(-\tilx^{2})$.}
\end{figure}

We investigate the traveling wave solutions of the SRL phase
transformation model by numerically solving \eqref{eqn:nondimbarc} with
an explicit finite difference method, second order in space and time.
Representative phase transformation waves during Li insertion and
extraction are shown in Fig.~\ref{fig:twave}.  For these simulations,
$\tila = 5$ and $\tillambda = 1$, similar to the values given in
\cite{han04}.  The choice of $\tillambda$ corresponds to the dimensional
lengthscales $\lambda = L = 10$ nm.  Although there is no available
experimental or simulational data on the rate coefficients $\kins$ and
$\kext$, we expect that the insertion and extraction reactions occur on
the same timescale and therefore assume $\kappa = 1$.  Insertion or
extraction is forced by raising or lowering the electrolyte chemical
potential $\tilmu_{e}$ to promote transfer in one direction and inhibit
transfer in the other; $\tilmu_{e} = 0.5$ for the insertion process in
Fig.~\ref{fig:twave}a, and $\tilmu_{e} = -1$ for the extraction process
in Fig.~\ref{fig:twave}b.  The initial conditions for the insertion and
extraction waves, denoted by dashed lines in the plots, are Gaussian
fluctuations of the composition representing nucleations of the
lithiated and unlithiated phases, respectively.  For the insertion in
Fig.~\ref{fig:twave}a, $\bar{c}(\tilx, 0) = 0.1 + 0.8 \exp(-\tilx^{2})$,
and for the extraction in Fig.~\ref{fig:twave}b, $\bar{c}(\tilx, 0) =
0.9 - 0.8 \exp(-\tilx^{2})$.

The initial composition fluctuation rapidly develops into two wavefronts
bounded by the equilibrium Li-poor and Li-rich phases $g_{1}$ and
$g_{3}$, respectively.  The development of a fully formed wavefront
involves both insertion and extraction.  During the insertion process in
Fig.~\ref{fig:twave}a, the maximum concentration of the initial
fluctuation grows to $g_{3}$, while the low concentration baseline
decays to $g_{1}$.  Conversely, for the extraction process in
Fig.~\ref{fig:twave}b, the minimum concentration decays to $g_{1}$, and
the high concentration baseline grows to $g_{3}$.  Once fully developed
wavefronts form, they propagate to the right and left with a constant
velocity.  We have verified that the velocity of a fully developed
wavefront is constant for all times in the numerical simulations, as
expected from the phase plane analysis described earlier.

\begin{figure}
\includegraphics[scale=1]{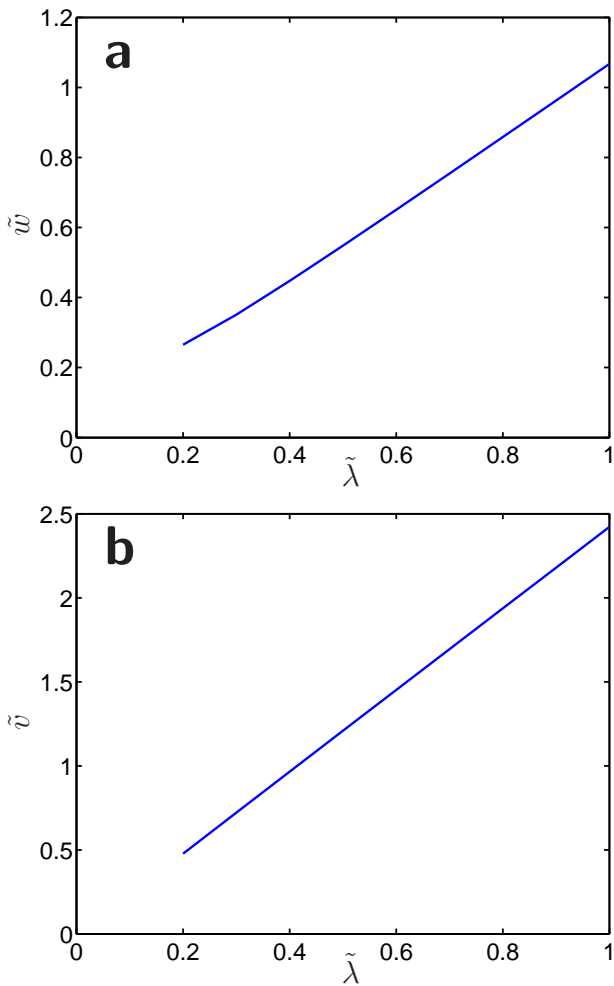}
\caption{\label{fig:waveproperties} Scaling of computed wave (a) width
and (b) velocity with $\tillambda$.  Parameters: $\tila = 5$, $\kappa =
1$, $\tilmu_{e} = 0.5$, $\bar{c}(\tilx, 0) = 0.1 + 0.8
\exp(-\tilx^{2})$.}
\end{figure}

The dimensionless width $\tilw$ and velocity $\tilv$ of a fully
developed wavefront are determined by the parameters of the system.  The
main parameter controlling the width of the wave is $\tillambda$, since
it contains the gradient energy coefficient $K$.  As $\tillambda$ is
decreased, the energetic penalty for forming gradients in the
concentration is lowered, resulting in sharper wavefronts spanning the
equilibrium phase compositions.  Additionally, we find that sharper
wavefronts move at a slower velocity.  The numerically computed
dependence of the width and velocity on $\tillambda$ is shown in
Fig.~\ref{fig:waveproperties}.  It is apparent that both $\tilw$ and
$\tilv$ scale linearly with $\tillambda$ in the range of physical
relevance.  In fact, this linear scaling can be determined analytically
by a dimensional analysis of the system in the sharp interface limit
$\tillambda \ll 1$, as performed in the following section.

\begin{figure}
\includegraphics[scale=1]{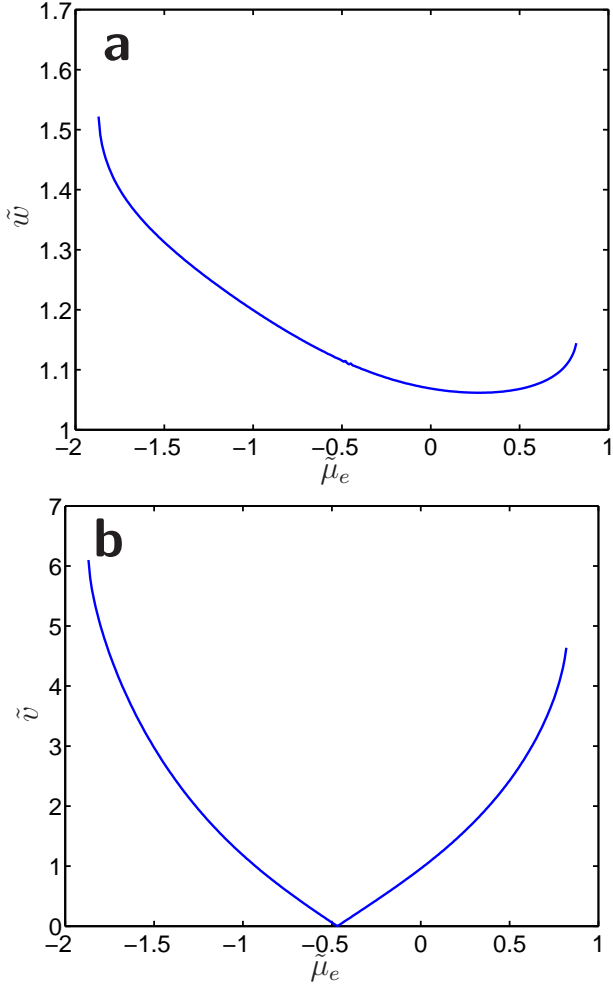}
\caption{\label{fig:muescaling} Dependence of computed wave (a) width
and (b) velocity on $\tilmu_{e}$.  Parameters: $\tila = 5$, $\kappa =
1$, $\tillambda = 1$, $\bar{c}(\tilx, 0) = g_{1}(\tilmu_{e}) + (1/2)
(g_{3}(\tilmu_{e}) - g_{1}(\tilmu_{e})) (\tanh(\tilx) + 1)$.}
\end{figure}

The electrolyte chemical potential $\tilmu_{e}$ is an experimentally
accessible independent parameter, as it corresponds to the applied
potential of the system, forcing Li insertion or extraction to occur.
Thus, it is important to consider the dependence of the wave width and
velocity on this parameter, as a systematic study of this dependence may
eventually lead to an experimentally feasible method of testing the SRL
model.  Fig.~\ref{fig:muescaling} shows the numerically determined
dependence of $\tilw$ and $\tilv$ on $\tilmu_{e}$.  As shown in
Fig.~\ref{fig:muescaling}a, the wave width exhibits a weak, nonlinear
dependence on $\tilmu_{e}$, embodied in the scaling function $F_{w}$
that arises in the dimensional analysis described in the following
section.  The velocity dependence in Fig.~\ref{fig:muescaling}b shows
how the system transitions from extraction waves to insertion waves as
$\tilmu_{e}$ increases from negative to positive.  Large negative values
of $\tilmu_{e}$ strongly force Li removal, resulting in extraction waves
with large velocities.  As $\tilmu_{e}$ increases, the extraction wave
velocity declines steeply, until zero velocity is obtained at
$\tilmu_{e} \approx -0.5$.  This point is the transition from extraction
to insertion.  Insertion waves with increasing velocities are produced
as $\tilmu_{e}$ increases beyond the transition point.  Analogously to
the width, the velocity dependence is contained within a scaling
function $F_{v}$ discussed in the following section.  Note that the
width and velocity profiles are not symmetric about the transition
point, and the minimum width does not correspond to zero velocity.  The
asymmetry in the width and velocity result from the asymmetry in the
homogeneous net reaction rate $\tilRinsH - \tilRextH$.  As described
previously, $\tilRinsH - \tilRextH$ must have three roots over the
composition range $0 < c < 1$ in order for phase separation to occur.
This requirement imposes a restricted range on $\tilmu_{e}$ for a given
set of parameters.  The extreme negative and positive values of
$\tilmu_{e}$ in Fig.~\ref{fig:muescaling} are therefore fixed; there are
no traveling wave solutions beyond them.  The bounding values of
$\tilmu_{e}$ can be determined analytically by solving
\eqref{eqn:gstnrypt} for the $\tilmu_{e}^{\mp}$ corresponding to the
extrema $g_{\pm}$ given by \eqref{eqn:gextrema}, as these compositions
define the limits of the phase separation range.  We obtain
\begin{equation}
\tilmu_{e}^{\mp} = \tila (1 - 2 g_{\pm}) + \ln \left(
\frac{g_{\pm}^{3/2}}{1 - g_{\pm}} \right) + \ln \sqrt{\kappa},
\end{equation}
where $\tilmu_{e}^{-}$ is the minimum allowable potential for extraction
waves, and $\tilmu_{e}^{+}$ is the maximum allowable potential for
insertion waves.  The notational $\pm$ signs of $g_{\pm}$ and
$\tilmu_{e}^{\mp}$ are reversed since $g_{-}$ is the minimum extremum at
which there is almost only insertion, hence corresponding to the maximum
allowable potential $\tilmu_{e}^{+}$, and conversely for $g_{+}$ and
$\tilmu_{e}^{-}$.  The limitations on $\tilmu_{e}$ are physically
meaningful.  In principle, transport can be driven by an arbitrarily
positive or negative $\tilmu_{e}$.  However, such a strong $\tilmu_{e}$
effectively increases the overall surface reaction rate $k$ and pushes
the system out of the SRL transport regime $Da \ll 1$.  Indeed, for
sufficiently fast surface reactions, the system becomes a BTL process
($Da \gg 1$) with phase transformation governed by shrinking core type
dynamics.

\begin{figure}
\includegraphics[scale=1]{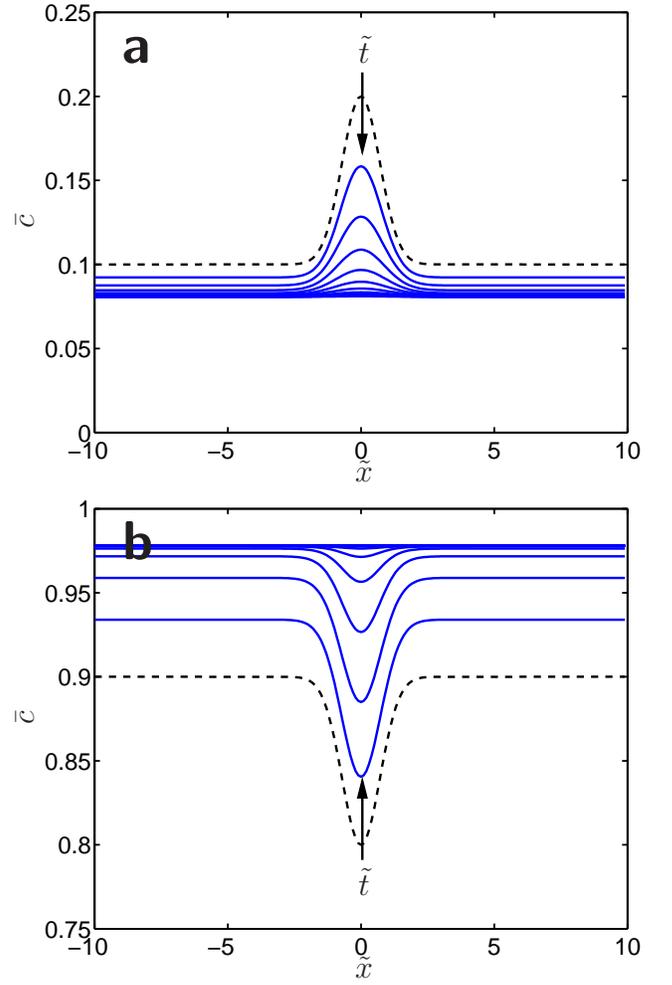}
\caption{\label{fig:notwave} Numerically computed failed insertion and
extraction events.  Dashed lines show initial conditions; solid lines
show concentration profiles at uniformly spaced times, with arrows
indicating direction of profile movement.  For both insertion and
extraction: $\tila = 5$, $\kappa = 1$, $\tillambda = 1$.  (a) Failed
insertion event with $\tilmu_{e} = 0.5$, $\bar{c}(\tilx, 0) = 0.1 + 0.1
\exp(-\tilx^{2})$.  (b) Failed extraction event with $\tilmu_{e} = -1$,
$\bar{c}(\tilx, 0) = 0.9 - 0.1 \exp(-\tilx^{2})$.}
\end{figure}

Not all initial conditions give rise to traveling waves, as shown in
Fig.~\ref{fig:notwave}.  The key requirement for the formation of
traveling waves is that the initial condition supports both addition and
removal of Li in the domain, that is $R(x, 0)$ must change sign.  The
simultaneous addition and removal of material sharpen the initial
composition fluctuation to a phase separating wavefront.  In the failed
insertion event shown in Fig.~\ref{fig:notwave}a, only extraction
occurs, and the initial composition perturbation decays to a uniform
concentration of $g_{1}$.  Similarly, Fig.~\ref{fig:notwave}b presents a
failed extraction event where the initial composition depression fills
up to a uniform concentration of $g_{3}$.  Note that $R(x, 0)$ depends
on the Laplacian of the initial concentration profile $\bar{c}(x, 0)$,
and thus different initial conditions with equal composition ranges but
varying spatial distributions will or will not produce traveling waves.
This behavior has been numerically verified.

The dependence of the formation of traveling waves on the initial
condition relates to the nucleation of the phase separation.  During
insertion, for example, the lithiated phase may first nucleate at some
atomic scale inhomogeneity on the crystal surface where it is
energetically favorable for Li atoms to collect.  Our theory does not
account for such features, and therefore it cannot model the initiation
of the phase separation.  Moreover, as the FePO$_{4}$/LiFePO$_{4}$ phase
boundary width $\lambda$ is nearly at the atomic scale, applying a
continuum equation for nucleation below this scale may not be physically
relevant.  We note, however, that continuum nucleation could be studied
in other SRL systems with larger $\lambda$.

\subsection{Scaling with material constants}

A complete characterization of wave dynamics in the SRL regime is beyond
the scope of this paper, but we conclude this section by summarizing
some key features from the analysis above, with dimensions restored.

\emph{Wave nucleation.}  Starting from a pure phase and adjusting the
external chemical potential in the electrolyte $\mu_{e}$, or the
interaction strength $a$, the nucleation of a phase transition wave
occurs whenever a random concentration fluctuation produces a region of
sufficiently high, stable concentration of the opposite phase. The
gradient penalty then sharpens the wave, and it propagates by the
addition or removal of ions at the wavefront, due it its elevated
chemical potential compared to both pure phases.

\emph{Wave width.}  By dimensional analysis, the width $w$ of the
traveling wave profile is given by
\begin{equation}
w = L F_{w} \left( \frac{\lambda}{L}, \frac{a}{\rho k T},
\frac{\mu_{e}}{\rho k T}, \frac{\kext}{\kins} \right)
\end{equation} 
where $F_{w}(\tillambda, \tila, \tilmu_e, \kappa)$ is a scaling
function.  Since the width $w$ should not depend on the size of the
crystal $L$ in the limit of a sharp phase boundary, we have $\tilw =
F_{w} \sim \tillambda f_{w}$ or $w \propto \lambda$ for $\tillambda \ll
1$, consistent with the numerical solutions above in
Fig.~\ref{fig:waveproperties}a.  With units restored, we see that the
width is set by the phase-boundary thickness
\begin{equation}
w \sim \sqrt{\frac{K}{\rho kT} } f_w  \left( \frac{a}{\rho k T},
\frac{\mu_{e}}{\rho k T}, \frac{\kext}{\kins} \right)
\end{equation}
as expected, although it may also depend weakly on $\tila$,
$\tilmu_{e}$, and $\kappa$.  A slice of the $\tilmu_{e}$ dependence is
shown in Fig.~\ref{fig:muescaling}a.

\emph{Wave speed.}  The speed of the traveling wave has a similar form,
\begin{equation}
v = \frac{L}{\tau} F_{v} \left( \frac{\lambda}{L}, \frac{a}{\rho k T},
\frac{\mu_{e}}{\rho k T}, \frac{\kext}{\kins}  \right)
\end{equation}
where $F_{v}(\tillambda, \tila, \tilmu_{e}, \kappa)$ is another scaling
function. Once again, the limit of a sharp phase boundary requires,
$\tilv = F_{v} \sim \tillambda f_{v}$ or $v \propto \lambda/\tau$ for
$\tillambda \ll 1$, consistent with the numerical solutions in
Fig.~\ref{fig:waveproperties}b.

Recalling the time unit, we obtain a general expression for the wave
speed in the limit of a sharp phase boundary,
\begin{equation}\label{eqn:vel}
  v \sim \sqrt{\frac{K}{\rho k T}} \frac{2 \rho_{s} \kins}{\rho L_{y}}
  f_{v} \left(\frac{a}{\rho k T}, \frac{\mu_{e}}{\rho k T},
    \frac{\kext}{\kins} \right).
\end{equation}
Note that the wave speed decreases with increasing crystal thickness, $v
\propto 1/L_{y}$ since it takes longer for reactions to fill each bulk
channel, as the SRL phase transformation proceeds sweeps across the
crystal. The speed is also proportional to the gradient penalty
coefficient $K$ and the insertion rate constant $\kins$. It also should
decrease with the strength of the interaction between ions in the
crystal $a$, which drives phase separation. The wave velocity can also
be controlled externally by varying the chemical potential of ions in
the electrolyte $\mu_{e}$, as shown in Fig.~\ref{fig:muescaling}b.

\section{Response to an applied voltage}

\subsection{BTL versus SRL dynamics}

Transport in electrode materials is often studied by measuring the
current response of the material to an applied potential.  In an
isotropic BTL process, a potential step induces a current proportional
to the diffusion limited flux of ions across the electrode/electrolyte
interface.  The response time of any linear or nonlinear diffusion
limited process, such as assumed in the shrinking core model, is given
by the characteristic time $t_{D}-L^2/D$.  The flux for small times, and hence
the current, can be found analytically from the similarity solution for
diffusion in a semi-infinite domain.  The resulting expression for the
current is known as the Cottrell equation,
\begin{equation}
I_{\text{Cottrell}} = n e A \rho \sqrt{\frac{D}{\pi t}}, \quad t
\ll t_{D},
\end{equation}
where $n$ is the number of electrons transferred and $A$ is the
electrode particle area.  The Cottrell current response forms the basis
of the Potentiostatic Intermittent Titration Technique (PITT) that is
commonly used to measure the diffusivity of materials \cite{wen79}.
However, cathodes that operate in a SRL transport regime where bulk
diffusion in a preferred direction is fast relative to ion transfer at
the electrode/electrolyte interface cannot be assumed to follow Cottrell
dynamics.

In our model for single crystal LiFePO$_{4}$, the chemical potential
of the electrolyte $\mu_{e}$ serves as the applied potential to the
system.  For an appropriate $\mu_{e}$, sharply defined waves of Li
propagate across the crystal surface as Li transfer occurs.  A
composition fluctuation initiates each wave, and all fully developed
waves in the system travel with the same, constant velocity, in a flat
plate-like particle.  The total flux of ions across the two $xz$
surfaces of the crystal is determined by the integral of the net
insertion rate, so the current response of a single crystal is given by 
\begin{equation}
I = n e \iint\! \rho_s R\, dx\, dz,
\end{equation}
Note that since the reaction rate $R$ is zero at the equilibrium phase
compositions $g_{1}$ and $g_{3}$, only localized wavefronts spanning
these compositions contribute to the current. 

The scaling of the SRL response is radically different from that of
the Cottrell BTL response. Ignoring geometrical effects and nucleation
(discussed below), the basic scaling of the response time for a single
crystal is given by
\begin{equation}
t_{v} = \frac{L}{v} = \frac{L \tau}{\lambda} = L \sqrt{\frac{\rho k
T}{K}} \frac{\rho L_{y}}{2 \rho_{s} \kins}.
\end{equation}
Note that the response time is proportional to two geometrical
lengths, the depth of the channels $L_{y}$ and the length $L$ over
which the waves are propagating. More importantly, the time is
determined by the surface reaction rate $\kins$ and not by the bulk
diffusion coefficient $D$. The ratio of the time scales for BTL
dynamics and SRL dynamics is an effective P\'eclet number, 
\begin{equation}
Pe = \frac{t_D}{t_v} = \frac{v L}{D} =  \sqrt{\frac{K}{\rho k
T}} \frac{2 \rho_{s} \kins L}{\rho L_{y} D}
\end{equation}
which measures the importance of wave propagation at the diffusive
time scale. However, this is once again the Damkohler number, since
the reaction time is set by wave propagation, $t_R = t_v$, in the SRL
regime.

\subsection{Plate-like crystals}

\begin{figure}
\includegraphics[scale=1]{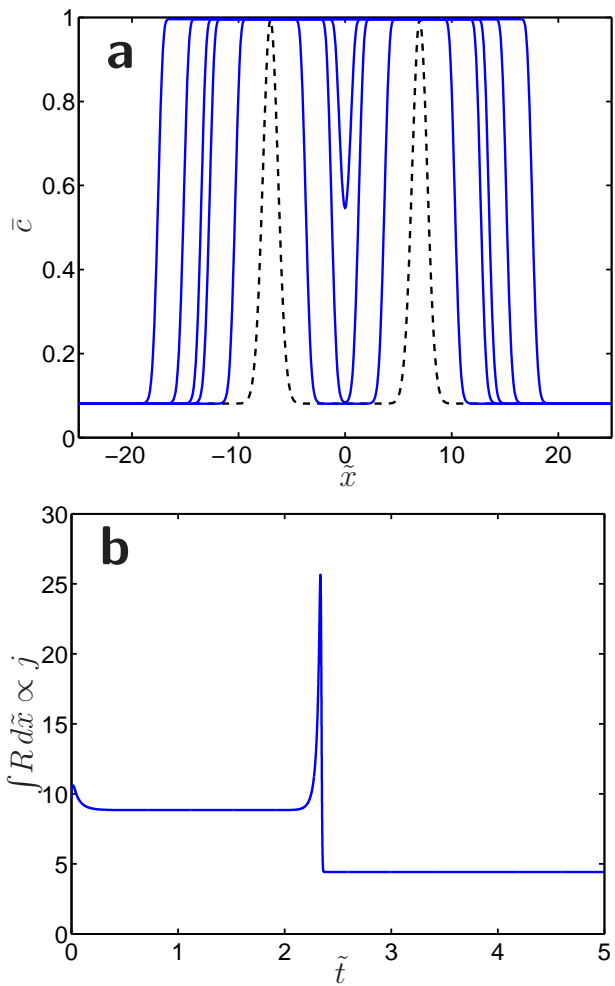}
\caption{\label{fig:twowavej} Numerically computed phase transformation
dynamics and flux response of two impinging insertion waves.  (a)
Concentration profiles of impinging waves.  Dashed line shows initial
condition; solid lines show concentration profile at various times
during impingement.  (b) Flux response of impinging waves.  Parameters:
$\tila = 5$, $\kappa = 1$, $\tillambda = 1$, $\tilmu_{e} = 0.5$,
$\bar{c}(\tilx, 0) = g_{1} + (g_{3} - g_{1}) [\exp(-(\tilx + 7)^{2}) +
\exp(-(\tilx - 7)^{2})]$.}
\end{figure}

To develop a general picture of SRL phase transformation dynamics in a
rechargeable battery cathode, we first consider the case of flat
plate-like crystals of constant depth $L_{y}$ analyzed in the previous
section. A fully developed wavefront moving with constant velocity
supports a steady current, and there is a sudden loss in the current
when two wavefronts merge and are replaced by an equilibrium
composition.  Fig.~\ref{fig:twowavej} shows this declining staircase
form for the current in a system with two impinging waves.  The spike
in the current at the time of collision is due to the gradient penalty
term in the reaction rate acting on the sharp composition profile of
the merging waves.  We note that the magnitude of the spike is large
in this simulation since there are only two waves; in an actual system
with many waves, the surge in current from any individual collision
would be small relative to the total current being sustained.

Thus, the current response of LiFePO$_{4}$ is governed by the overall
rate of its transformation through concurrent nucleation and growth of
waves.  Phase nucleation likely occurs by both heterogeneous and
homogeneous mechanisms.  Recent first principles computations have found
that the chemical potential of Li varies considerably over the surface
of the equilibrium crystal shape \cite{wang07}.  Consequently, different
crystal faces may be energetically favorable for heterogeneous phase
nucleation during Li insertion and extraction.

\begin{figure}
\includegraphics[scale=1]{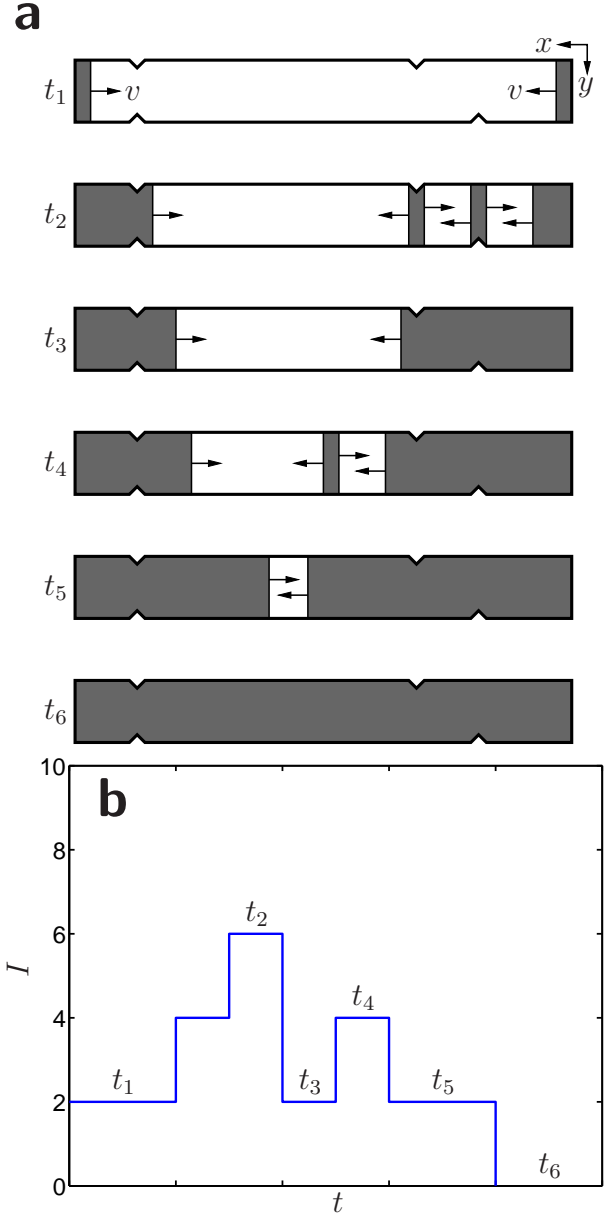}
\caption{\label{fig:nucandgrow} Schematic diagram of overall phase
transformation and current response of a single plate-like crystal
undergoing SRL transport.  (a) Sequence of $xy$ cross sections of
crystal at times $t_{1}, \ldots, t_{6}$, illustrating phase
transformation of material through concurrent nucleation and growth of
traveling waves.  Each fully developed wavefront moves with velocity $v$
and sustains a constant current of unity.  (b) Declining staircase
current response of crystal, with labeled times corresponding to
illustrations in (a).}
\end{figure}

An example of a single crystal undergoing heterogeneous and homogeneous
nucleation and growth is illustrated in Fig.~\ref{fig:nucandgrow}.
Fig.~\ref{fig:nucandgrow}a shows $xy$ cross sections of the crystal at a
sequence of successive times $t_{1}, \ldots, t_{6}$, and
Fig.~\ref{fig:nucandgrow}b presents the corresponding profile of the
total current $I$.  At time $t_{1}$, heterogeneous nucleation at the
crystal edges has produced two fully developed wavefronts, each moving
with constant velocity $v$ and sustaining a normalized current of unity.
Therefore, the crystal supports the total current $I = 2$ at this time.
At some time between $t_{1}$ and $t_{2}$, heterogeneous nucleation
occurs at the two rightmost surface defects of the crystal, indicated by
notches.  Once these nucleation events grow into fully developed waves,
there are 6 propagating wavefronts carrying a total current of $I = 6$.
The rightmost waves merge at time $t_{3}$ such that 4 wavefronts are
destroyed, and consequently, only two traveling wavefronts remain and
the current drops to $I = 2$.  Homogeneous nucleation at some location
in the untransformed fraction of the material occurs at time $t_{4}$,
and the two additional wavefronts created increase the current to $I =
4$.  The rightmost waves combine at time $t_{5}$, and as most of the
material is transformed and only two wavefronts remain, the current
again drops to $I = 2$.  Finally, at time $t_{6}$, the material is fully
transformed and can no longer sustain a current.

\subsection{Other crystal shapes}

The wave dynamics can depend sensitively on the crystal \emph{shape} in
the SRL regime. This general fact can be easily seen from our analysis
of a plate-like crystal: the wave velocity \eqref{eqn:vel} depends on
the local depth $L_{y}(x,z)$ of the fast diffusion channels in the bulk,
as well as the local surface orientation, through the surface-site
density $\rho_{s}(x,z)$. The analysis of SRL phase-transformation
dynamics for arbitrary crystal shapes is a challenging problem, left for
future work.

Here, we simply indicate how scalings by considering the limit of slowly
varying depth, $L_{y}(x)$, and assuming the 1D wave dynamics for a flat
surface are only slightly perturbed. The local wave velocity is then
\begin{equation}\label{eqn:vode}
v = \frac{dx_w}{dt} = \frac{D_w}{L_y(x_w)}
\end{equation}
where $D_{w} = 2 \rho_{s}(x_{w}) \kins \sqrt{K/(\rho^{3} k T)} \approx$
constant.  This ordinary differential equation can be solved for the
position of the wave $x_{w}(t)$ from a given wave nucleation event for
any slowly varying shape $L_{y}(x_{w})$. For example, consider the case
of a cylinder, $L_{y}(x) = \sqrt{R^{2} - x^{2}}$ (ignoring that the
shape is not slowly varying at the ends). In that case, the equation can
be solved analytically, e.g., for nucleation at one edge $x = -R$. For
early times, the wave velocity decays from its initial value as
$dx_{w}/dt \propto D_{w}^{2/3}t^{-1/3}$, due to the increasing depth of
the bulk channels. 

In spite of variable wave speed, however, the current remains constant
during wave propagation in this approximation, as in the case of a flat
plate, regardless of the crystal shape. The reason is that wave
propagation at speed $v(x_{w})$ engulfs channels of length
$L_{y}(x_{w})$, so the total current, proportional to $v L_{y}$, remains
constant according to \eqref{eqn:vode}. The time for a wave to engulf
the entire crystal thus scales with the total volume (in contrast to
diffusive BTL dynamics, where the time scales like the cross-sectional
area).  Physically, ions are being inserted or extracted at roughly a
constant rate, since the phase boundary is assumed to have a constant
exposed length at the surface for 1D dynamics.

For spheres and other 3D shapes, the wavefront will not remain flat, and
the full 2D depth-integrated dynamics will need to be solved with
variable $L_{y}(x,z)$ and $\rho_{s}(x,z)$. However, the current may
remains roughly constant during wave propagation, since the filling time
for a channel is proportional to its length, at constant surface
reaction rate. In that case, the results of the previous section for
current versus time in flat plate-like single crystals may not be
substantially modified with more complicated shapes, although
statistical fluctuations due to random nucleation events will be
different. 

\subsection{Composite cathode response}

It is important to note that Fig.~\ref{fig:nucandgrow} represents only
one possible realization of the transformation and current response of
the crystal.  Heterogeneous nucleation may occur at different edges or
surface defects at different times for different insertion and
extraction cycles.  Homogeneous nucleation would be spatially
distributed in some random fashion.  Therefore, to determine the overall
current response of a composite cathode composed of many individual
crystals, we must consider the statistical distributions of the
nucleation events.

For homogeneous nucleation, we may assume that the nucleation rate is
uniform across the crystal surface.  Nucleation events in the
untransformed material are independent, and the presence of a previously
nucleated wave does not influence the likelihood of nucleation around
that wave.  With these assumptions, the nucleation events are
distributed as a Poisson process in time, and we may invoke the
Johnson-Mehl-Avrami equation for the overall transformation rate of the
material \cite{ballufi05}
\begin{equation}
\xi = e^{-G t^{n}},
\end{equation}
where $\xi$ is the untransformed fraction of the material, $G$ is a
factor dependent on the dimensionality of the process, and the integer
$n > 1$; for a 1D line nucleation process, $G t^{n} \sim t^{2}$.  The
Johnson-Mehl-Avrami equation therefore specifies that the overall
transformation rate of the material follows a sigmoidal shape.  We thus
expect that the current also follows this sigmoidal response for
concurrent homogeneous nucleation and growth.

In the case of heterogeneous nucleation, we may assume that for many
defects in many particles, the heterogeneous nucleation events are
distributed as a Poisson process in space.  Thus, the qualitative form
of the average response is expected to follow the same sigmoidal shape
as homogeneous nucleation, assuming all electrode particles are at the
same potential (driving force), which may only be true under low rate
conditions or for thin electrodes \cite{delacourt05, gaberscek07}.

\begin{figure}
\includegraphics[scale=1]{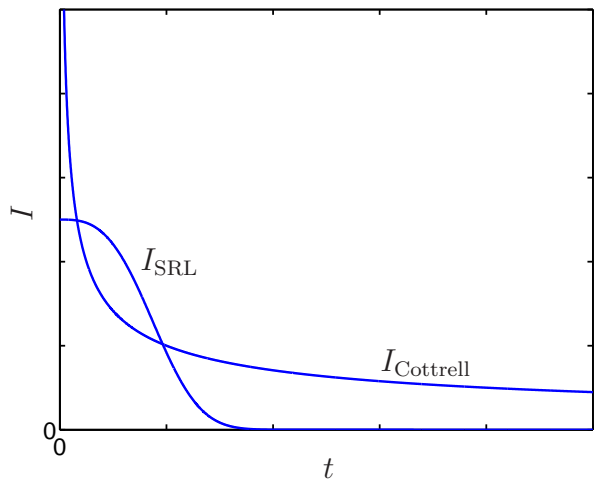}
\caption{\label{fig:cottrellandsigmoid} Schematic illustration of
composite cathode responses for Cottrell (BTL) and SRL phase
transformation mechanisms.}
\end{figure}

Thus, we have found that individual crystals have a declining staircase
form for their current response to an applied electrolyte potential
$\mu_{e}$ if a substantial number of nucleation events occur on the
timescale of complete transformation.  For many crystals, we may
consider that the homogeneous and heterogeneous nucleation events are
Poisson processes in time and space, respectively.  The overall
transformation and current response of the material is then given by the
average of many distinct staircase responses, resulting in a
characteristic sigmoidal curve
\begin{equation}
I \sim e^{-G t^{n}},
\end{equation}
that is strikingly different than the Cottrell response that is commonly
assumed.  Fig.~\ref{fig:cottrellandsigmoid} compares the sigmoidal and
Cottrell responses.  We conclude that PITT measurements in material
undergoing SRL transport do not measure the diffusivity.  Rather, these
measurements provide some measure of the kinetic parameters in the
surface reaction rates that are controlling the overall rate of the
material transformation.

\subsection{Discussion}

The model we propose is based on a few key experimental and theoretical
findings.  (1) Li migrates rapidly in the (010) direction ($y$ in
Fig.~\ref{fig:crystal}) creating either filled or empty channels that
completely penetrate the material.  This makes it possible to
coarse-grain to a two-dimensional model in which the surface
concentration defines the concentration in a channel.  (2) A linear
interface exists on the particle surface between filled and unfilled
sites, and growth of one phase at the expense of the other occurs by
displacement of this interface.  Such one-dimensional growth is
supported by experimental observations \cite{chen06} and by a recent
Johnson-Mehl Avrami analysis of the growth exponents \cite{allen07}.
Since one would expect two-dimensional growth on the surface for an
isotropic material, the one-dimensional growth has to find its origin in
the crystallography of the material.  Undoubtedly, it is either the
anisotropy of the interfacial strain energy - because of different
coherence strains or elastic constants \cite{maxisch06b}, or the
anisotropy of the interfacial energy which causes the system to prefer a
single interface plane.

In our model, the transformation rate is controlled by transfer of the
ions from the electrolyte to the (010) surface.  Since during growth
insertion only takes place at the interface, the energy of the Li ions
at this interface is a key quantity in determining the effective
transfer rate from the solution.  Unlike in a core-shell model where the
ability of the particle to take up current declines as (de)lithiation
proceeds, in our model the transformation rate is nominally constant,
until the particle is either fully transformed, new nucleation events
occur, or two wave fronts impinge.  Such transformation kinetics is
fundamentally different from Cottrell-like behavior.

How this transformation kinetics manifests itself in the observable
voltage-current response may depend very much on the structure of the
macroscopic electrode in which the active LiFePO$_{4}$ particles are
embedded.  In addition to the active material, a typical electrode
contains about 5--10 wt\% polymeric binder and 5--15 wt\% carbon black
to enhance electronic conductivity through the electrode. Some porosity
is also created in the electrode to allow the electrolyte to penetrate
and transport the Li$^{+}$ ions to and from the active material.  If the
conductive pathways for the Li$^{+}$ ions and electrons are sufficient,
all particles will be at the same potential and experience a similar
driving force for transformation.  Under these conditions, and assuming
stochastic nucleation, we expect that the overall current response to a
potential pulse is sigmoidal.  Recent work indicates that such
equipotential conditions across the electrode only apply at rather low
charge and discharge rates, or for very thin electrodes
\cite{gaberscek07}.  If such electrical resistance along the thickness
of the electrode plays a role, the collective current response of the
system could be viewed as of a sum of sigmoidals, each with a different
driving force, but time-dependent screening effects would also need to
be taken into account.

\section{\label{sec:conclusion}Conclusion}

We have proposed a general continuum theory for Li transport in single
crystal LiFePO$_{4}$ based on a thermodynamically sound phase field
formulation of the free energy.  In various limits of the characteristic
timescales for bulk and surface transport, this theory captures the
shrinking core and other BTL models.  In the limit where fast bulk
diffusion in 1D equilibrates the bulk concentration to the surface
concentration, our model gives a new type of equation describing SRL
phase transition dynamics.  We find that this model exhibits traveling
wave solutions that qualitatively agree with the experimental
observations.  The main implication of our work for battery modeling is
that the current response of SRL systems, such as LiFePO$_{4}$, are not
governed by the classical, bulk diffusion limited Cottrell model that is
commonly assumed.  Our work also focuses attention to the importance of
the Li$^{+}$ and electron delivery to the proper surface of LiFePO$_{4}$
in order to achieve fast charge absorption.  While much effort in the
experimental literature has focused on electron delivery (e.g. by carbon
coating or conductive Fe$_{2}$P contributions) \cite{ravet01, herle04},
little emphasis seems to have been place on rapid transport of Li$^{+}$
towards the surface where it can penetrate.  Finally, we note that the
SRL model developed here may be applicable in other materials where
surface transfer effects are rate limiting, such as nanoporous
materials.

\begin{acknowledgments}
This work was supported by the MRSEC program of the National Science
Foundation under award number DMR 02-12383. The authors thank
K.~Thornton for helpful comments on the manuscript. 
\end{acknowledgments}

\bibliography{LiFePO4}

\begin{thebibliography}{35}
\expandafter\ifx\csname natexlab\endcsname\relax\def\natexlab#1{#1}\fi
\expandafter\ifx\csname bibnamefont\endcsname\relax
  \def\bibnamefont#1{#1}\fi
\expandafter\ifx\csname bibfnamefont\endcsname\relax
  \def\bibfnamefont#1{#1}\fi
\expandafter\ifx\csname citenamefont\endcsname\relax
  \def\citenamefont#1{#1}\fi
\expandafter\ifx\csname url\endcsname\relax
  \def\url#1{\texttt{#1}}\fi
\expandafter\ifx\csname urlprefix\endcsname\relax\def\urlprefix{URL }\fi
\providecommand{\bibinfo}[2]{#2}
\providecommand{\eprint}[2][]{\url{#2}}

\bibitem[{\citenamefont{Padhi et~al.}(1997)\citenamefont{Padhi, Nanjundaswamy,
  and Goodenough}}]{padhi97}
\bibinfo{author}{\bibfnamefont{A.~K.} \bibnamefont{Padhi}},
  \bibinfo{author}{\bibfnamefont{K.~S.} \bibnamefont{Nanjundaswamy}},
  \bibnamefont{and} \bibinfo{author}{\bibfnamefont{J.~B.}
  \bibnamefont{Goodenough}}, \bibinfo{journal}{J. Electrochem. Soc.}
  \textbf{\bibinfo{volume}{144}}, \bibinfo{pages}{1188} (\bibinfo{year}{1997}).

\bibitem[{\citenamefont{Yamada et~al.}(2001)\citenamefont{Yamada, Chung, and
  Hinokuma}}]{yamada01}
\bibinfo{author}{\bibfnamefont{A.}~\bibnamefont{Yamada}},
  \bibinfo{author}{\bibfnamefont{S.~C.} \bibnamefont{Chung}}, \bibnamefont{and}
  \bibinfo{author}{\bibfnamefont{K.}~\bibnamefont{Hinokuma}},
  \bibinfo{journal}{J. Electrochem. Soc.} \textbf{\bibinfo{volume}{148}},
  \bibinfo{pages}{A224} (\bibinfo{year}{2001}).

\bibitem[{\citenamefont{Chung et~al.}(2002)\citenamefont{Chung, Bloking, and
  Chiang}}]{chung02}
\bibinfo{author}{\bibfnamefont{S.~Y.} \bibnamefont{Chung}},
  \bibinfo{author}{\bibfnamefont{J.~T.} \bibnamefont{Bloking}},
  \bibnamefont{and} \bibinfo{author}{\bibfnamefont{Y.~M.}
  \bibnamefont{Chiang}}, \bibinfo{journal}{Nat. Mater.}
  \textbf{\bibinfo{volume}{1}}, \bibinfo{pages}{123} (\bibinfo{year}{2002}).

\bibitem[{\citenamefont{Delacourt et~al.}(2005)\citenamefont{Delacourt, Poizot,
  Tarascon, and Masquelier}}]{delacourt05}
\bibinfo{author}{\bibfnamefont{C.}~\bibnamefont{Delacourt}},
  \bibinfo{author}{\bibfnamefont{P.}~\bibnamefont{Poizot}},
  \bibinfo{author}{\bibfnamefont{J.~M.} \bibnamefont{Tarascon}},
  \bibnamefont{and}
  \bibinfo{author}{\bibfnamefont{C.}~\bibnamefont{Masquelier}},
  \bibinfo{journal}{Nat. Mater.} \textbf{\bibinfo{volume}{4}},
  \bibinfo{pages}{254} (\bibinfo{year}{2005}).

\bibitem[{\citenamefont{Morgan et~al.}(2004)\citenamefont{Morgan, der Ven, and
  Ceder}}]{morgan04}
\bibinfo{author}{\bibfnamefont{D.}~\bibnamefont{Morgan}},
  \bibinfo{author}{\bibfnamefont{A.~V.} \bibnamefont{der Ven}},
  \bibnamefont{and} \bibinfo{author}{\bibfnamefont{G.}~\bibnamefont{Ceder}},
  \bibinfo{journal}{Electrochem. Solid State Lett.}
  \textbf{\bibinfo{volume}{7}}, \bibinfo{pages}{A30} (\bibinfo{year}{2004}).

\bibitem[{\citenamefont{Ouyang et~al.}(2004)\citenamefont{Ouyang, Shi, Wang,
  Huang, and Chen}}]{ouyang04}
\bibinfo{author}{\bibfnamefont{C.~Y.} \bibnamefont{Ouyang}},
  \bibinfo{author}{\bibfnamefont{S.~Q.} \bibnamefont{Shi}},
  \bibinfo{author}{\bibfnamefont{Z.~X.} \bibnamefont{Wang}},
  \bibinfo{author}{\bibfnamefont{X.~J.} \bibnamefont{Huang}}, \bibnamefont{and}
  \bibinfo{author}{\bibfnamefont{L.~Q.} \bibnamefont{Chen}},
  \bibinfo{journal}{Phys. Rev. B} \textbf{\bibinfo{volume}{69}},
  \bibinfo{pages}{104303} (\bibinfo{year}{2004}).

\bibitem[{\citenamefont{Islam et~al.}(2005)\citenamefont{Islam, Driscoll,
  Fisher, and Slater}}]{islam05}
\bibinfo{author}{\bibfnamefont{M.~S.} \bibnamefont{Islam}},
  \bibinfo{author}{\bibfnamefont{D.~J.} \bibnamefont{Driscoll}},
  \bibinfo{author}{\bibfnamefont{C.~A.~J.} \bibnamefont{Fisher}},
  \bibnamefont{and} \bibinfo{author}{\bibfnamefont{P.~R.}
  \bibnamefont{Slater}}, \bibinfo{journal}{Chem. Mat.}
  \textbf{\bibinfo{volume}{17}}, \bibinfo{pages}{5085} (\bibinfo{year}{2005}).

\bibitem[{\citenamefont{Zhou et~al.}(2006)\citenamefont{Zhou, Maxisch, and
  Ceder}}]{zhou06}
\bibinfo{author}{\bibfnamefont{F.}~\bibnamefont{Zhou}},
  \bibinfo{author}{\bibfnamefont{T.}~\bibnamefont{Maxisch}}, \bibnamefont{and}
  \bibinfo{author}{\bibfnamefont{G.}~\bibnamefont{Ceder}},
  \bibinfo{journal}{Phys. Rev. Lett.} \textbf{\bibinfo{volume}{97}},
  \bibinfo{pages}{155704} (\bibinfo{year}{2006}).

\bibitem[{\citenamefont{Maxisch and Ceder}(2006)}]{maxisch06b}
\bibinfo{author}{\bibfnamefont{T.}~\bibnamefont{Maxisch}} \bibnamefont{and}
  \bibinfo{author}{\bibfnamefont{G.}~\bibnamefont{Ceder}},
  \bibinfo{journal}{Phys. Rev. B} \textbf{\bibinfo{volume}{73}},
  \bibinfo{pages}{174112} (\bibinfo{year}{2006}).

\bibitem[{\citenamefont{Xu et~al.}(2004)\citenamefont{Xu, Chung, Bloking,
  Chiang, and Ching}}]{xu04}
\bibinfo{author}{\bibfnamefont{Y.~N.} \bibnamefont{Xu}},
  \bibinfo{author}{\bibfnamefont{S.~Y.} \bibnamefont{Chung}},
  \bibinfo{author}{\bibfnamefont{J.~T.} \bibnamefont{Bloking}},
  \bibinfo{author}{\bibfnamefont{Y.~M.} \bibnamefont{Chiang}},
  \bibnamefont{and} \bibinfo{author}{\bibfnamefont{W.~Y.} \bibnamefont{Ching}},
  \bibinfo{journal}{Electrochem. Solid State Lett.}
  \textbf{\bibinfo{volume}{7}}, \bibinfo{pages}{A131} (\bibinfo{year}{2004}).

\bibitem[{\citenamefont{Zhou et~al.}(2004)\citenamefont{Zhou, Kang, Maxisch,
  Ceder, and Morgan}}]{zhou04}
\bibinfo{author}{\bibfnamefont{F.}~\bibnamefont{Zhou}},
  \bibinfo{author}{\bibfnamefont{K.~S.} \bibnamefont{Kang}},
  \bibinfo{author}{\bibfnamefont{T.}~\bibnamefont{Maxisch}},
  \bibinfo{author}{\bibfnamefont{G.}~\bibnamefont{Ceder}}, \bibnamefont{and}
  \bibinfo{author}{\bibfnamefont{D.}~\bibnamefont{Morgan}},
  \bibinfo{journal}{Solid State Commun.} \textbf{\bibinfo{volume}{132}},
  \bibinfo{pages}{181} (\bibinfo{year}{2004}).

\bibitem[{\citenamefont{Maxisch et~al.}(2006)\citenamefont{Maxisch, Zhou, and
  Ceder}}]{maxisch06a}
\bibinfo{author}{\bibfnamefont{T.}~\bibnamefont{Maxisch}},
  \bibinfo{author}{\bibfnamefont{F.}~\bibnamefont{Zhou}}, \bibnamefont{and}
  \bibinfo{author}{\bibfnamefont{G.}~\bibnamefont{Ceder}},
  \bibinfo{journal}{Phys. Rev. B} \textbf{\bibinfo{volume}{73}},
  \bibinfo{pages}{104301} (\bibinfo{year}{2006}).

\bibitem[{\citenamefont{Chen et~al.}(2006)\citenamefont{Chen, Song, and
  Richardson}}]{chen06}
\bibinfo{author}{\bibfnamefont{G.~Y.} \bibnamefont{Chen}},
  \bibinfo{author}{\bibfnamefont{X.~Y.} \bibnamefont{Song}}, \bibnamefont{and}
  \bibinfo{author}{\bibfnamefont{T.~J.} \bibnamefont{Richardson}},
  \bibinfo{journal}{Electrochem. Solid State Lett.}
  \textbf{\bibinfo{volume}{9}}, \bibinfo{pages}{A295} (\bibinfo{year}{2006}).

\bibitem[{\citenamefont{Laffont et~al.}(2006)\citenamefont{Laffont, Delacourt,
  Gibot, Wu, Kooyman, Masquelier, and Tarascon}}]{laffont06}
\bibinfo{author}{\bibfnamefont{L.}~\bibnamefont{Laffont}},
  \bibinfo{author}{\bibfnamefont{C.}~\bibnamefont{Delacourt}},
  \bibinfo{author}{\bibfnamefont{P.}~\bibnamefont{Gibot}},
  \bibinfo{author}{\bibfnamefont{M.~Y.} \bibnamefont{Wu}},
  \bibinfo{author}{\bibfnamefont{P.}~\bibnamefont{Kooyman}},
  \bibinfo{author}{\bibfnamefont{C.}~\bibnamefont{Masquelier}},
  \bibnamefont{and} \bibinfo{author}{\bibfnamefont{J.~M.}
  \bibnamefont{Tarascon}}, \bibinfo{journal}{Chem. Mat.}
  \textbf{\bibinfo{volume}{18}}, \bibinfo{pages}{5520} (\bibinfo{year}{2006}).

\bibitem[{\citenamefont{Amin et~al.}(2007)\citenamefont{Amin, Balaya, and
  Maier}}]{amin07}
\bibinfo{author}{\bibfnamefont{R.}~\bibnamefont{Amin}},
  \bibinfo{author}{\bibfnamefont{P.}~\bibnamefont{Balaya}}, \bibnamefont{and}
  \bibinfo{author}{\bibfnamefont{J.}~\bibnamefont{Maier}},
  \bibinfo{journal}{Electrochem. Solid State Lett.}
  \textbf{\bibinfo{volume}{10}}, \bibinfo{pages}{A13} (\bibinfo{year}{2007}).

\bibitem[{\citenamefont{Meethong et~al.}(2007)\citenamefont{Meethong, Huang,
  Speakman, Carter, and Chiang}}]{meethong07}
\bibinfo{author}{\bibfnamefont{N.}~\bibnamefont{Meethong}},
  \bibinfo{author}{\bibfnamefont{H.~Y.~S.} \bibnamefont{Huang}},
  \bibinfo{author}{\bibfnamefont{S.~A.} \bibnamefont{Speakman}},
  \bibinfo{author}{\bibfnamefont{W.~C.} \bibnamefont{Carter}},
  \bibnamefont{and} \bibinfo{author}{\bibfnamefont{Y.~M.}
  \bibnamefont{Chiang}}, \bibinfo{journal}{Adv. Funct. Mater.}
  \textbf{\bibinfo{volume}{17}}, \bibinfo{pages}{1115} (\bibinfo{year}{2007}).

\bibitem[{\citenamefont{Srinivasan and Newman}(2004)}]{srinivasan04}
\bibinfo{author}{\bibfnamefont{V.}~\bibnamefont{Srinivasan}} \bibnamefont{and}
  \bibinfo{author}{\bibfnamefont{J.}~\bibnamefont{Newman}},
  \bibinfo{journal}{J. Electrochem. Soc.} \textbf{\bibinfo{volume}{151}},
  \bibinfo{pages}{A1517} (\bibinfo{year}{2004}).

\bibitem[{\citenamefont{Doyle et~al.}(1993)\citenamefont{Doyle, Fuller, and
  Newman}}]{doyle93}
\bibinfo{author}{\bibfnamefont{M.}~\bibnamefont{Doyle}},
  \bibinfo{author}{\bibfnamefont{T.~F.} \bibnamefont{Fuller}},
  \bibnamefont{and} \bibinfo{author}{\bibfnamefont{J.}~\bibnamefont{Newman}},
  \bibinfo{journal}{J. Electrochem. Soc.} \textbf{\bibinfo{volume}{140}},
  \bibinfo{pages}{1526} (\bibinfo{year}{1993}).

\bibitem[{\citenamefont{Han et~al.}(2004)\citenamefont{Han, der Ven, Morgan,
  and Ceder}}]{han04}
\bibinfo{author}{\bibfnamefont{B.~C.} \bibnamefont{Han}},
  \bibinfo{author}{\bibfnamefont{A.~V.} \bibnamefont{der Ven}},
  \bibinfo{author}{\bibfnamefont{D.}~\bibnamefont{Morgan}}, \bibnamefont{and}
  \bibinfo{author}{\bibfnamefont{G.}~\bibnamefont{Ceder}},
  \bibinfo{journal}{Electrochim. Acta} \textbf{\bibinfo{volume}{49}},
  \bibinfo{pages}{4691} (\bibinfo{year}{2004}).

\bibitem[{\citenamefont{y~de Dompablo et~al.}(2002)\citenamefont{y~de Dompablo,
  der Ven, and Ceder}}]{dedompablo02}
\bibinfo{author}{\bibfnamefont{M.~E.~A.} \bibnamefont{y~de Dompablo}},
  \bibinfo{author}{\bibfnamefont{A.~V.} \bibnamefont{der Ven}},
  \bibnamefont{and} \bibinfo{author}{\bibfnamefont{G.}~\bibnamefont{Ceder}},
  \bibinfo{journal}{Phys. Rev. B} \textbf{\bibinfo{volume}{66}},
  \bibinfo{pages}{064112} (\bibinfo{year}{2002}).

\bibitem[{\citenamefont{der Ven and Ceder}(2004)}]{vanderven04}
\bibinfo{author}{\bibfnamefont{A.~V.} \bibnamefont{der Ven}} \bibnamefont{and}
  \bibinfo{author}{\bibfnamefont{G.}~\bibnamefont{Ceder}},
  \bibinfo{journal}{Electrochem. Commun.} \textbf{\bibinfo{volume}{6}},
  \bibinfo{pages}{1045} (\bibinfo{year}{2004}).

\bibitem[{\citenamefont{Cahn and Hilliard}(1958)}]{cahn58}
\bibinfo{author}{\bibfnamefont{J.~W.} \bibnamefont{Cahn}} \bibnamefont{and}
  \bibinfo{author}{\bibfnamefont{J.~E.} \bibnamefont{Hilliard}},
  \bibinfo{journal}{J. Chem. Phys.} \textbf{\bibinfo{volume}{28}},
  \bibinfo{pages}{258} (\bibinfo{year}{1958}).

\bibitem[{\citenamefont{Khachaturyan}(1983)}]{khachaturyan83}
\bibinfo{author}{\bibfnamefont{A.~G.} \bibnamefont{Khachaturyan}},
  \emph{\bibinfo{title}{Theory of Structural Transformations in Solids}}
  (\bibinfo{publisher}{Wiley-Interscience}, \bibinfo{address}{New York},
  \bibinfo{year}{1983}).

\bibitem[{\citenamefont{Larche and Cahn}(1985)}]{larche85}
\bibinfo{author}{\bibfnamefont{F.~C.} \bibnamefont{Larche}} \bibnamefont{and}
  \bibinfo{author}{\bibfnamefont{J.~W.} \bibnamefont{Cahn}},
  \bibinfo{journal}{Acta. Met.} \textbf{\bibinfo{volume}{33}},
  \bibinfo{pages}{331} (\bibinfo{year}{1985}).

\bibitem[{\citenamefont{der Ven and Ceder}(2000)}]{vanderven00}
\bibinfo{author}{\bibfnamefont{A.~V.} \bibnamefont{der Ven}} \bibnamefont{and}
  \bibinfo{author}{\bibfnamefont{G.}~\bibnamefont{Ceder}},
  \bibinfo{journal}{Electrochem. Solid State Lett.}
  \textbf{\bibinfo{volume}{3}}, \bibinfo{pages}{301} (\bibinfo{year}{2000}).

\bibitem[{\citenamefont{Wang et~al.}(2007)\citenamefont{Wang, Zhou, Meng, and
  Ceder}}]{wang07}
\bibinfo{author}{\bibfnamefont{L.}~\bibnamefont{Wang}},
  \bibinfo{author}{\bibfnamefont{F.}~\bibnamefont{Zhou}},
  \bibinfo{author}{\bibfnamefont{Y.~S.} \bibnamefont{Meng}}, \bibnamefont{and}
  \bibinfo{author}{\bibfnamefont{G.}~\bibnamefont{Ceder}},
  \bibinfo{journal}{submitted}  (\bibinfo{year}{2007}).

\bibitem[{\citenamefont{Murray}(2002)}]{murray02}
\bibinfo{author}{\bibfnamefont{J.~D.} \bibnamefont{Murray}},
  \emph{\bibinfo{title}{Mathematical Biology I}}
  (\bibinfo{publisher}{Springer-Verlag}, \bibinfo{address}{New York},
  \bibinfo{year}{2002}).

\bibitem[{\citenamefont{Volpert et~al.}(1994)\citenamefont{Volpert, Volpert,
  and Volpert}}]{volpert94}
\bibinfo{author}{\bibfnamefont{A.~I.} \bibnamefont{Volpert}},
  \bibinfo{author}{\bibfnamefont{V.~A.} \bibnamefont{Volpert}},
  \bibnamefont{and} \bibinfo{author}{\bibfnamefont{V.~A.}
  \bibnamefont{Volpert}}, \emph{\bibinfo{title}{Traveling Wave Solutions of
  Parabolic Systems}} (\bibinfo{publisher}{American Mathematical Society},
  \bibinfo{address}{Providence, Rhode Island}, \bibinfo{year}{1994}).

\bibitem[{\citenamefont{Brunet and Derrida}(1997)}]{brunet97}
\bibinfo{author}{\bibfnamefont{E.}~\bibnamefont{Brunet}} \bibnamefont{and}
  \bibinfo{author}{\bibfnamefont{B.}~\bibnamefont{Derrida}},
  \bibinfo{journal}{Phys. Rev. E} \textbf{\bibinfo{volume}{56}},
  \bibinfo{pages}{2597} (\bibinfo{year}{1997}).

\bibitem[{\citenamefont{Wen et~al.}(1979)\citenamefont{Wen, Boukamp, Huggins,
  and Weppner}}]{wen79}
\bibinfo{author}{\bibfnamefont{C.~J.} \bibnamefont{Wen}},
  \bibinfo{author}{\bibfnamefont{B.~A.} \bibnamefont{Boukamp}},
  \bibinfo{author}{\bibfnamefont{R.~A.} \bibnamefont{Huggins}},
  \bibnamefont{and} \bibinfo{author}{\bibfnamefont{W.}~\bibnamefont{Weppner}},
  \bibinfo{journal}{J. Electrochem. Soc.} \textbf{\bibinfo{volume}{126}},
  \bibinfo{pages}{2258} (\bibinfo{year}{1979}).

\bibitem[{\citenamefont{Ballufi et~al.}(2005)\citenamefont{Ballufi, Allen, and
  Carter}}]{ballufi05}
\bibinfo{author}{\bibfnamefont{R.~W.} \bibnamefont{Ballufi}},
  \bibinfo{author}{\bibfnamefont{S.~M.} \bibnamefont{Allen}}, \bibnamefont{and}
  \bibinfo{author}{\bibfnamefont{W.~C.} \bibnamefont{Carter}},
  \emph{\bibinfo{title}{Kinetics of Materials}} (\bibinfo{publisher}{John Wiley
  \& Sons}, \bibinfo{address}{Hoboken, New Jersey}, \bibinfo{year}{2005}).

\bibitem[{\citenamefont{Gaberscek et~al.}(2007)\citenamefont{Gaberscek, Kuzma,
  and Jamnik}}]{gaberscek07}
\bibinfo{author}{\bibfnamefont{M.}~\bibnamefont{Gaberscek}},
  \bibinfo{author}{\bibfnamefont{M.}~\bibnamefont{Kuzma}}, \bibnamefont{and}
  \bibinfo{author}{\bibfnamefont{J.}~\bibnamefont{Jamnik}},
  \bibinfo{journal}{Phys. Chem. Chem. Phys.} \textbf{\bibinfo{volume}{9}},
  \bibinfo{pages}{1815} (\bibinfo{year}{2007}).

\bibitem[{\citenamefont{Allen et~al.}(2007)\citenamefont{Allen, Jow, and
  Wolfenstine}}]{allen07}
\bibinfo{author}{\bibfnamefont{J.~L.} \bibnamefont{Allen}},
  \bibinfo{author}{\bibfnamefont{T.~R.} \bibnamefont{Jow}}, \bibnamefont{and}
  \bibinfo{author}{\bibfnamefont{J.}~\bibnamefont{Wolfenstine}},
  \bibinfo{journal}{Chem. Mat.} \textbf{\bibinfo{volume}{19}},
  \bibinfo{pages}{2108} (\bibinfo{year}{2007}).

\bibitem[{\citenamefont{Ravet et~al.}(2001)\citenamefont{Ravet, Chouinard,
  Magnan, Besner, Gauthier, and Armand}}]{ravet01}
\bibinfo{author}{\bibfnamefont{N.}~\bibnamefont{Ravet}},
  \bibinfo{author}{\bibfnamefont{Y.}~\bibnamefont{Chouinard}},
  \bibinfo{author}{\bibfnamefont{J.~F.} \bibnamefont{Magnan}},
  \bibinfo{author}{\bibfnamefont{S.}~\bibnamefont{Besner}},
  \bibinfo{author}{\bibfnamefont{M.}~\bibnamefont{Gauthier}}, \bibnamefont{and}
  \bibinfo{author}{\bibfnamefont{M.}~\bibnamefont{Armand}},
  \bibinfo{journal}{J. Power Sources} \textbf{\bibinfo{volume}{97-8}},
  \bibinfo{pages}{503} (\bibinfo{year}{2001}).

\bibitem[{\citenamefont{Herle et~al.}(2004)\citenamefont{Herle, Ellis, Coombs,
  and Nazar}}]{herle04}
\bibinfo{author}{\bibfnamefont{P.~S.} \bibnamefont{Herle}},
  \bibinfo{author}{\bibfnamefont{B.}~\bibnamefont{Ellis}},
  \bibinfo{author}{\bibfnamefont{N.}~\bibnamefont{Coombs}}, \bibnamefont{and}
  \bibinfo{author}{\bibfnamefont{L.~F.} \bibnamefont{Nazar}},
  \bibinfo{journal}{Nat. Mater.} \textbf{\bibinfo{volume}{3}},
  \bibinfo{pages}{147} (\bibinfo{year}{2004}).

\end{thebibliography}

\end{document}